\documentclass[12pt, a4paper]{article}
\usepackage{graphicx} 
\usepackage{parskip} 
\usepackage{amsmath}
\usepackage{braket} 
\usepackage{enumitem}
\usepackage{amssymb}
\usepackage{amsthm}
\usepackage{slashed}
\usepackage{wrapfig}
\usepackage{mathtools}
\usepackage[backend=biber,style=nature,doi=true,url=true]{biblatex}
\usepackage{xurl}
\usepackage[hidelinks]{hyperref}

\newcommand\blfootnote[1]{%
  \begingroup
  \renewcommand\thefootnote{}\footnote{#1}%
  \addtocounter{footnote}{-1}%
  \endgroup
}

\title{The Birth of Quantum Mechanics: A Historical Study Through the Canonical Papers}
\author{
    \small Eren Volkan Küçük \\
    \small Department of Physics and Astronomy, Universität Heidelberg \\
    \small Heidelberg, Germany \\
    \small \texttt{eren.kucuk@stud.uni-heidelberg.de}
    }

\date{}

\begin{document}
\setlength{\baselineskip}{1.25\baselineskip}
\setlength{\parskip}{1.25em}

\maketitle

\begin{abstract}
    This paper explores the historical development of the theory of quantum mechanics between 1900 and 1927 by chronological examination of the foundational papers and ideas. Beginning with Planck's introduction of energy quantisation in blackbody radiation, we follow the emergence of Einstein's light quanta hypothesis, Bohr's atomic model, and the statistical implications of indistinguishable particles. Special emphasis is placed on the transition from the Old Quantum Theory to modern quantum mechanics, particularly through Heisenberg's matrix mechanics and Schrödinger's wave mechanics. This study aims to provide a structured historical account, offering insights into the conceptual transformations that led to quantum mechanics while making the development accessible to physicists, historians of science, and advanced students interested in the origins of modern quantum theory.
    
\end{abstract}

\section{Introduction}
The development of quantum physics stands as one of the most profound revolutions in the history of science. Before the advent of quantum theory, classical physics, governed by Newtonian mechanics and Maxwell's electrodynamics, was considered to provide a complete description of nature. However, towards the late 19th and early 20th centuries, several inconsistencies emerged that could not be resolved within the classical framework. One of the most significant issues was black body radiation, where classical predictions failed to match experimental observations \cite{Kirchhoff1860, Wien1896, Rayleigh1900, ehrenfest1911}. Another problem was the photoelectric effect, a detailed experimental study of which had been done by Philip Lenard \cite{lenard1902}, which challenged the classical wave theory of light. After Rutherford suggested his atom model \cite{rutherford1911}, the stability of atoms also posed a major dilemma since, according to classical electrodynamics, accelerating electrons in orbit around the nucleus should continuously emit radiation and spiral into the nucleus, contradicting the observed stability of matter. Another problem was that the spectral lines in atomic emission and absorption spectra exhibited some patterns that classical mechanics could not explain (Jammer, 1989, ch.2.1). These anomalies indicated a fundamental flaw in classical theories, necessitating a new framework for understanding microscopic phenomena. These were, in fact, the cries of a newly arriving revolution in our understanding of the universe. This paper explores the historical development of these theoretical concepts, tracing the key ideas and discoveries that led to the formulation of quantum mechanics until the development of wave mechanics and uncertainty principle.

For a physicist, a thorough understanding of the thought processes of the minds who put these concepts forward is instrumental, if not a must, in seeing its problems and scope and ultimately improving further on these considerations. Therefore, the aim of this work is to present a concise history of the development of the theory of quantum physics through the study of the "canonical" papers to varying degrees. This study roughly follows the structure of Max Jammer's seminal work, \textit{The Conceptual Development of Quantum Mechanics} \cite{Jammer1966}, which provides a comprehensive historical analysis on the subject matter. Being an immensely detailed and profound work, which also gets into the details of many experimental studies\footnote{Since the goal of this study is to understand the development of theoretical ideas rather than experiments (although the significance of which for the formation of these ideas are not ignored) the relevant experiments were not studied in detail.} and some additional philosophical discussions related to the field in comparison to the present work, it is highly recommended for further study.

Of course, for a work of this scope, the omission of numerous significant works was unavoidable. Throughout the work, we put special emphasis on the papers of Heisenberg, Born, Jordan, Dirac, de Broglie, and Schrödinger. We tried to be as concise as possible and sometimes let the calculations speak for themselves whenever the results were clear. The calculations are presented as concisely as possible by a selection and summarisation of the ideas and computations from the relevant pages. Since this criterion of "relevance" is attributed to the papers and then to the parts of the papers somewhat subjectively by the author, they are, of course, open to criticism.

\section{The Black Body Radiation}
The birth of quantum theory is generally considered as Planck's papers on black body radiation \cite{Planck1901, Planck1900}.  There were two equations for such an object's radiation's energy density: The Rayleigh-Jeans law \cite{Rayleigh1900, Jeans1905} (cast in its usual form),
\begin{equation}
    u(f, T) = \frac{8 \pi}{c^2} f^2 kT,
\end{equation}
where $8 \pi f^2/c^2$ represents the density of modes\footnote{As usual $f$ is the frequency, and $c$ is the speed of light in vacuum, $k$ is the Boltzmann constant, and $T$ is the temperature in Kelvin scale.}; and Wien's law \cite{Wien1896},
\begin{equation}
    u(f, T) = \alpha f^3 e^{-\beta f/T },
\end{equation}
where $\alpha$ and $\beta$ are constants. These equations worked well for long and short wavelengths, respectively, but blew up in other regimes and didn't match up, where the former also foresaw the so-called UV catastrophe \cite{ehrenfest1911}.\footnote{The form of the equation put forward by Rayleigh in \cite{Rayleigh1900}, with the exponential cutoff inserted at the end, did not pose such a problem. But Rayleigh thought of the expression as an approximate result, which is in part theoretical and in part empirical, that worked for large wavelengths, as he says in \cite{Rayleigh1900}:
\begin{quote}
...The suggestion is that (4) rather than, as according to (2),
\begin{equation*}
    \lambda^{-5} \: d\lambda
\end{equation*}
may be the proper form when $\lambda \theta$ [$\theta$ is the temperature] is great. If we introduce the exponential factor, the complete expression will be
\begin{equation*}
    c_1 \theta \lambda^{-4} e^{-c_2/\lambda \theta} d\lambda.
\end{equation*} 
\end{quote}
In his 1905 paper \cite{Einstein1905}, which is the topic of the next section, Einstein stated that the classical theory inevitably leads to the Rayleigh-Jeans law (without the empirically inserted cutoff) and showed that it blows up in high frequencies. For more information, see \cite{Bomark2023}.} The main problem was that according to classical electromagnetism and statistical mechanics, a black body, which is an idealised perfect absorber of any frequency of electromagnetic radiation in thermal equilibrium (which can theoretically be represented by radiation in a cavity with perfectly reflecting walls with a hole \cite{Kirchhoff1860}), the total energy of radiation was infinite according to the Rayleigh-Jeans law. Since each distinct frequency mode of the oscillators in a black body may radiate in totally arbitrary amplitudes, they should have had equal average energies according to classical electrodynamics. The reason for that was the equipartition theorem, which attained an equal amount of energy on average for every degree of freedom of the electromagnetic field. So, for infinitely many frequency modes $f$, which all have an equal amount of average energy, gave rise to an infinite total energy.

In 1900, Planck proposed a new expression that fit the data better \cite{Planck1900}. In the following paper \cite{Planck1901}, he used statistical mechanics and classical electromagnetism to derive the formula more rigorously, and it was in that paper that he proposed the quantisation of energy states. Planck solved the problem by introducing energy quantisation for the field and leaving out the equipartition theorem. Thereby, oscillations of high frequency had higher energy so that they were harder to excite or, equivalently, far less improbable for the field to shine in. It is important to observe that Planck introduced energy quantisation to be able to perform the calculation, which was concerned with statistical entropy and thus with the possible arrangments of the energies of the oscillators. That, in turn, worked as a cut-off for high frequencies, and UV catastrophe was eliminated.

From statistical mechanics, the partition function for a single mode $f$ for the continuous energy levels is
\begin{equation}
    Z = \sum_n e^{-\beta E(f_n)} \approx \int_0^\infty e^{-\beta E} dE = 1/\beta,
\end{equation}
where $f_n$ represents the discrete frequency mode for the standing waves of the cavity, representing the black body (note that $\beta$ is not a constant this time). The average energy for a mode is,
\begin{equation}
    \overline{E} = -\frac{1}{Z}\frac{\partial Z}{\partial \beta} = 1/\beta = kT,
\end{equation}
(as equipartition of energy foresees) and since there are an infinite number of modes (frequencies), the total energy is infinite.

If one assumes that energy levels of a mode $f_n$ are dependent on the frequency as $E_{m,n} = mhf_n$ just as Planck did, where $m \in \mathbb{N}$, the partition function is
\begin{equation}
    Z = \sum_m e^{-\beta E_m(f_n)} = \sum_m e^{-\beta mhf_n} = \frac{1}{1 - e^{-\beta hf_n}}.
\end{equation}
In that case, for high $f$, the probability for higher energy states exponentially diminishes. Therefore, the average energy per mode is
\begin{equation}
    \overline{E} = -\frac{1}{Z}\frac{\partial Z}{\partial \beta} = \frac{hf_n}{e^{\beta hf_n} - 1}.
\end{equation}
Since the energy density of states is represented by $8\pi f^2/c^2$, the total energy is
\begin{equation}
    E_{tot} = \int_0^{\infty} \frac{8 \pi f^2}{c^2} \overline{E}(f, T) df = \frac{8 \pi}{c^2} \int_0^\infty  \frac{hf^3}{e^{\beta hf} - 1} df = \sigma T^4
\end{equation}
which is the Stefan-Boltzmann law.

One can observe that Rayleigh-Jeans and Wien's laws are short and long-wavelength regimes approximations of Planck's law. For low frequencies $hf/kT) << 1$,
\begin{equation}
    e^{\beta hf} - 1 \approx 1 + hf/kT) - 1 \Longrightarrow u = \frac{8 \pi}{c^2} f^2 kT;
\end{equation}
whereas for $hf/kT) >> 1$,
\begin{equation}
    u(f, T) = \frac{8 \pi f^2}{c^2} hf e^{-f/kT)},
\end{equation}
where $\alpha =  8 \pi h/c^2$, and $\beta = 1/k$.

\section{Einstein's Paper on Radiation Quanta}
The work of Planck marked the entry of quantum concepts into mainstream physics. The next step of great significance was Einstein's theory of radiation quanta and the explanation of the photoelectric effect \cite{Einstein1905}, published in the same year as his two other significant works on Brownian motion\footnote{This work of Einstein, although not directly related to the development of quantum mechanics, since posed a proof of the existence of atoms, it seemed relevant to study it in an appendix. Therefore, for more information on the paper, see the first appendix.}, and special relativity. The main problem which led Einstein to the following considerations was the apparent inconsistency of the concepts of discreteness of energy states, whereas the radiation itself was still considered continuous through classical electrodynamics. He also showed that the classical considerations inevitably led to the Rayleigh-Jeans law, without the exponential cutoff, which resulted in the UV-catastrophe presented in the previous section.

Einstein started his contemplations by stating that the entropy of the radiation in the cavity can be represented as a function of the radiation density, $u_\nu$, and $\nu$:
\begin{equation}
    S = V\int_0^\infty \varphi(u, \nu) d\nu,
\end{equation}
where $V$ is the volume of the cavity. Considering that energy is conserved and entropy is at its maximum,
\begin{equation}
    \begin{split}
        & \delta \int \varphi d\nu = 0, \\
        & \delta \int u d\nu = 0.
    \end{split}
\end{equation}
Therefore,
\begin{equation}
    \int \frac{\partial \varphi}{\partial u}\delta u d \nu = \lambda \int \delta u d\nu \Longrightarrow \int \left(\frac{\partial \varphi}{\partial u} - \lambda \right) \delta u d \nu = 0.
\end{equation}
Since this equality has to hold for any variation of $u$,
\begin{equation}
    \frac{\partial \varphi}{\partial u} = \lambda,
\end{equation}
which means, $\frac{\partial \varphi}{\partial u}$ is independent of $\nu$. The differential entropy is
\begin{equation}
    S = \int_0^\infty \frac{\partial \varphi}{\partial u} du d\nu,
\end{equation}
for $V = 1$. Since the derivative is independent of $\nu$, as we have shown, we have
\begin{equation}
    dS = \frac{\partial \varphi}{\partial u} dE = T^{-1} dE,
\end{equation}
where the last inequality is valid for reversible systems.

Wien's radiation law, which is valid for large values of $\nu$, states
\begin{equation}
    u_\nu = \alpha \nu^3 e^{-\beta \frac{\nu}{T}} \Longrightarrow T^{-1} = \partial_u \varphi =  -\frac{1}{\beta \nu }\ln{\frac{\rho}{\alpha \nu^3}}.
\end{equation}
By direct integration, we get
\begin{equation}
    \varphi = -\frac{\rho}{\beta \nu} \left( \ln{\frac{\rho}{\alpha \nu^3}} - 1 \right).
\end{equation}
For a radiation in volume $V$ and frequency range $\nu + d\nu$,
\begin{equation}
    S = V \varphi d\nu = -\frac{V \rho d\nu}{\beta \nu} \left( \ln{\frac{V \rho d\nu}{V \alpha \nu^3 d\nu}} - 1 \right).
\end{equation}
Since the energy density scales as $\rho \propto 1/V$, energy $V\rho d\nu$ stays constant if we change the volume reversibly to $V_0$. The entropy difference in this case is
\begin{equation}
\label{s-s0 1}
    S - S_0 = -\frac{E}{\beta \nu} \left( \ln{\frac{E}{V \alpha \nu^3 d\nu}} - 1 \right) + \frac{E}{\beta \nu} \left( \ln{\frac{E}{V_0 \alpha \nu^3 d\nu}} + 1 \right) = \frac{E}{\beta \nu} \ln{\frac{V_0}{V}}.
\end{equation}
This is just the entropy change for an ideal gas, which is the first sign that radiation behaves like particles in this case. Afterwards, Einstein this time used the statistical model of entropy (before which he used the thermodynamical one),
\begin{equation}
    S = k \ln W.
\end{equation}
He thought a volume of $n$ particles that don't interact with each other (so that "the radiation may be considered as the most disordered process imaginable"). For the same process, the entropy difference gives
\begin{equation}
    S - S_0 = k \ln{ \left( \frac{V}{V_0} \right)^n} = nk \ln{ \left( \frac{V}{V_0} \right)}.
\end{equation}
Comparing this to \eqref{s-s0 1}, it can be seen that we must have
\begin{equation}
    nk = n \frac{R}{N} = \frac{E}{\beta \nu} \Longrightarrow = E/n = \frac{R\beta \nu}{N}, 
\end{equation}
which is the energy per radiation quanta of frequency $\nu$ within the range of validity of Wien's radiation formula. Afterwards, he showed that the photoelectric effect is explained by these results.

Lastly, we have to mention a nuance: Notice that what Planck suggested is the quantisation of energy, whereas Einstein's is the quantisation of electromagnetic waves directly. Einstein thought that classical electrodynamics was an approximation of many light quanta and sufficiently long time intervals. He saw the irreconcilability of Planck's law, which he took for granted, and classical electrodynamics, whereas Planck did not accept the idea of quanta of radiation. Planck viewed quantisation as a mathematical convenience, whereas Einstein argued that light itself is quantised, laying the foundation for quantum mechanics. Therefore, the concept of quanta became a topic of discussion firstly as a mathematical convenience and/or as the atom of light.

\section{Emergence of the Old Quantum Theory}
Both Einstein's and Planck's papers stimulated a discussion in the physics community. The concept of quanta has taken place in many papers in the succeeding years, some of which are chosen to be discussed in this section.

In 1911, Ehrenfest and Kamerlingh considered the statistical mechanics of quantised radiation, showing that the statistical behaviour of distinguishable quanta leads to the Rayleigh-Jeans formula, while indistinguishable quanta result in the form of Planck's radiation law. They showed that for $N$ unit volumes and $P$ quanta, if the quanta are distinguishable as in Einstein's conception, we have
\begin{equation}
    W_1/W_2 = N_1^P/N_2^P,
\end{equation}
which leads to the Rayleigh-Jeans formula; whereas if the quanta are indistinguishable, as in Planck's theory\footnote{Of course, although Planck did introduce the concept of quantised energy levels, he did not initially interpret radiation itself as being composed of quanta. Later, the papers of Bose and Einstein showed that treating radiation as a gas of indistinguishable quanta (photons) naturally leads to the same distribution that Planck had derived through different means \cite{Einstein_1924, einstein1925}.}, we have
\begin{equation}
    W_1/W_2 = \begin{pmatrix}
        N_1 - 1 + P \\
        P
    \end{pmatrix}/\begin{pmatrix}
        N_2 - 1 + P \\
        P
    \end{pmatrix},
\end{equation}
which leads to Planck's law of radiation \cite{KamerlinghOnnesEhrenfest1914}. Today, we understand that the discrepancy arises because photons, like all bosons, are indistinguishable particles. Thus, Planck's formula correctly accounts for their quantum statistics. However, at the time, the concept of indistinguishable quantum particles had not yet been fully developed. Although their work did not yet formulate Bose-Einstein statistics, it foreshadowed the necessity of treating quantum particles as indistinguishable; an idea later formalised by Bose and Einstein in the 1920s \cite{einstein1925}.

The same year, in the first Solvay Conference, Planck contributed with a paper \cite{Planck1913} in which he proposed the idea that $h$ was the quantum of action. The consideration is as follows: Planck started with the classical theory of Gibbs, that the probability of finding the system in the state corresponding to the element $dpdq$ in phase space was
\begin{equation}
    \frac{e^{-\beta E} dpdq }{\int e^{-\beta E} dpdq }.
\end{equation}
For the harmonic oscillator, the calculation yields the famous result $kT$. Planck, then, stated that if we postulate that $E$ can only take on discrete values $E =  n \epsilon$, the calculation yields
\begin{equation}
    U = \frac{\epsilon}{e^{\beta \epsilon} -1  }.
\end{equation}
Therefore, Planck made the statement that not only the average energy of the oscillators is determined with respect to the preceding result, but directly the energy itself of the oscillators at any moment can take on the energy values according to it. According to these considerations, Planck proposed that $h$ is the finite extension of the elementary area in phase space. Therefore, the subsequent ellipses in the phase space of the harmonic oscillator (the trajectories) have a difference of area
\begin{equation}
    \int_{E}^{E + \epsilon} dpdq = h,
\end{equation}
or equivalently, all ellipses enclose an area of
\begin{equation}
    \int dpdq = nh.
\end{equation}
Therefore, he rendered the idea of the "quanta of the phase space" more fundamental than the idea of energy quanta, which he then considered to be only a consequence of the postulate. Thus, the problems of quantum mechanics were to be solved by first solving the problem classically and then applying the quantum of action condition. This postulate was to become one of the cornerstones of the Old Quantum Theory and, as we will see, be included implicitly in the foundations of matrix mechanics. It is also interesting to note that Planck started his account by stating that the framework of classical mechanics is too narrow to take into account the phenomena that are outside of our "coarse senses" .\footnote{From \cite{Planck1913}, p.93: "Nous devons reconnaître aujourd'hui que cette espérance n’était
pas justifiée et que le cadre de la Dynamique classique, même en
tenant compte de l’extension apportée par le principe de relativité
de Lorentz-Einstein, est trop étroit pour contenir les phénomènes
physiques non directement accessibles à nos moyens grossiers de
perception."} Here, we observe that the concept of quanta has taken a more fundamental place than it had before.

In the subsequent year, Poincare proved that the only way to achieve a radiation law of finite total energy was through the concept of quanta \cite{poincare1921}. The general idea is summarised as follows (Prentis, 1995, p.340):
\begin{quote}
    The global structure of the memoir is as follows. Poincaré
begins with a physical model consisting of a system of Hertzian resonators interacting with a system of atoms. The atoms, by virtue of their collisions with the resonators, mediate
the exchange of energy between the resonators. For both
physical reasons and mathematical rigour, Poincare felt that it
was necessary to clearly understand how a unique thermodynamic equilibrium state could emerge from an explicit consideration of some mechanism of interaction between the
resonators. He felt that the other theoretical studies of blackbody radiation were flawed because they neglected such an
interaction, and their arguments included too many assumptions related to the existence of equilibrium. He focuses on
the partition of energy between the resonators and the atoms
in thermal equilibrium. Poincaré's clever choice to study the
partition of energy allows him to avoid the usual arguments
based on entropy, temperature, and other equilibrium concepts. From his formulation of statistical mechanics, Poincaré is able to relate this macroscopic partition of energy to
the microscopic dynamical behaviour of a resonator. Having
established this basic micro-macro connection, he then determines what kind of constraint must be imposed on the
motion (energetics) of the resonators in order to account for
the observed partition of energy (radiation law). In three
separate proofs, Poincaré discovers that this dynamical constraint is precisely the hypothesis of quanta.
\end{quote}

\subsection{Bohr's Quantization of States in Atoms}
The next year, Niels Bohr, at the time still a young researcher, published a paper which was concerned primarily with the structure and spectrum of the hydrogen atom. According to the results of Rutherford \cite{rutherford1911}, electrons were scattered around the positively charged nucleus.\footnote{It turns out that Rutherford did not really suggest an "atom model", yet merely stated that there is a positively charged nucleus, without really touching upon how electrons move about it. For a discussion on the misconception on the atom models, see the discussion of Renstrøm and Bomark (2022) \textit{[arXiv:2212.08572]}.} Yet the laws of classical electrodynamics dictate that in such a system, electrons should radiate all their energy while spiralling down into the nucleus, wherefore there is no classical stable solution for such a system. To get rid of this problem, and taking inspiration from the regularities in the hydrogen spectrum which is made explicit by Balmer's formula, he applied Planck's quantization to the orbits and postulated that there are stationary states around the nucleus \cite{bohr1913}.\footnote{Before him, Planck's postulate had been applied to the oscillation of the electrons in the atom model of J.J. Thomson. See \cite{Jammer1966}, p.75.} In these discrete orbitals, the motion of the electron was governed by classical physics, but the transitions between the states were not, but occur as jumps between the discrete orbits. The theory was successful in explaining the spectra of the hydrogen atom, and mono-electronic ions in general, but the theory struggled to account for more complex systems. 

One of the most consulted principles, the correspondence principle as it is called later, in the Old Quantum Theory, which signifies the state of the theory before the next leap of understanding with primarily Heisenberg, de Broglie, and Schrödinger, was stated in the same paper. According to the results, the ratio of the frequency of revolution of the electron in subsequent orbits converges to $1$ as the orbital number increases. In his later paper in 1918, Bohr stated the principle more explicitly (Bohr, 1918):
\begin{quote}
    We shall
show, however, that the conditions which will be used to determine the
values of the energy in the stationary states are of such a type that the
frequencies calculated by (1), in the limit where the motions in successive
stationary states comparatively differ very little from each other, will tend
to coincide with the frequencies to be expected on the ordinary theory of
radiation from the motion of the system in the stationary states.\footnote{ The translation is from \cite{bohr1918}, p.10.}
\end{quote}
So, for large quantum numbers, the radiation is expected to obey the results of classical electrodynamics. Bohr's correspondence principle was crucial because it ensured that quantum mechanics would reproduce classical results in the limit of large quantum numbers, providing a bridge between old and new physics.

Later, Born suggested that the transition between classical to quantum mechanical equations can be achieved by replacing differentials with difference equations according to the rule \cite{born1924},
\begin{equation}
\label{correspondance principle}
    \tau \frac{\partial \Phi(n, \tau)}{\partial n} \leftrightarrow \Phi_{n + \tau, n} - \Phi_{n, n - \tau}.
\end{equation}
This follows from the following logic: The transition frequency from $n$ to $n' = n - \tau$ coincides with the $\tau^{th}$ harmonic of the classical motion in state $n$ for large $n$, that is
\begin{equation}
    \nu_{n,n-\tau} = \nu(n, \tau) = \tau \nu(n,1) = \tau \nu,
\end{equation}
where, in Jacobi-Hamilton formulation, in terms of action-angle variables,\footnote{See section B.2 in the appendix.} we also have
\begin{equation}
    \frac{\partial H}{\partial J} = \nu.
\end{equation}
Now, since $J = nh$, this equation becomes
\begin{equation}
    \nu(n, \tau) = \tau\frac{\partial H}{\partial J} = \frac{\tau}{h}\frac{\partial H}{\partial n} = \nu.
\end{equation}
Since we also have
\begin{equation}
    \nu_{n, n - \tau} = \frac{H(nh) - H([n - \tau]h)}{h},
\end{equation}
for an arbitrary function of stationary states $\Phi(n, \tau)$, we have the relation \eqref{correspondance principle}.  

\subsection{Sommerfeld's Elaborations of Bohr's and Planck's Results}
Later, in 1915, Sommerfeld derived the results of Bohr using the quantum of action for three degrees of freedom of the electron \cite{sommerfeld1915, sommerfeld1916}, for which the equations
\begin{equation}
    \oint p_r dr = n_rh, \: \oint p_\theta d\theta = n_\theta h, \: \oint p_\phi d\phi = n_\phi h,
\end{equation}
should hold. He studied the problem relativistically, and derived the relativistic correction to the spectrum of the hydrogen atom,
\begin{equation}
    E_{nk} = -Z^2Rhc \left[ \frac{1}{n^2} \frac{\alpha^2 Z^2}{n^4} \left( \frac{n}{k} - \frac{3}{4} \right)  \right],
\end{equation}
where $n = n_r + n_\phi $, $Z$ is the atom number, and $R$ is the Rydberg constant. This formula accounted for the energy splitting of the spectral lines and yielded the fine structure of the hydrogen atom, which is
\begin{equation}
    \alpha = \frac{2 \pi e^2}{hc}.
\end{equation}

Despite their partial success, these considerations did not save the quantum theory from its inconsistencies. As Jammer puts it, the theory at that state was "a lamentable hodgepodge of hypotheses, principles, theorems, and computational recipes rather than a logically consistent theory," (Jammer, 1966, p.208). The general method for solving a problem was first solving it classically, and then applying the "principles", as correspondance or quantum of action principle, applications of which most of the time asked for educated guesses. In short, the theory needed a logically consistent framework.

The lack of a consistent, logically sound framework in the Old Quantum Theory prompted Heisenberg to seek a new foundation for quantum mechanics, one based not on classical analogies (which were primarily concerned with unobservable properties of the systems such as position and velocity) but on directly observable quantities. Instead of these concepts, he insisted on placing at the core of the theory what he called the observable variables like energy and frequency, from which arise the so-called matrix mechanics \cite{Heisenberg1925}.

\section{Matrix Mechanics}
Throughout the paper, Heisenberg has in mind an electron in an atom which undergoes periodic motion. A time-dependent quantity $x_n(t)$, related to the electron in the $n^{th}$ state, can be expressed as a Fourier series in classical physics as
\begin{equation}
    x_n(t) = \sum_{\tau = -\infty}^{+\infty} x(n, \tau) = \sum_{\tau = -\infty}^{+\infty} x(n, \tau) \exp{[ 2\pi i \nu(n, \tau) t]},
\end{equation}
where $\tau ^{th}$ component has the frequency $\nu(n, \tau) = \tau \nu(n, 1)$. So it follows that
\begin{equation}
\label{classical frequency relation}
    \nu(n, \tau) + \nu(n, \tau') = \nu(n, \tau + \tau').
\end{equation}
The quantum mechanical analogue to this is
\begin{equation}
\label{quantum mechanical frequency sum rule}
    \nu_{n, n - \tau'} + \nu_{n - \tau', n - \tau} = \nu_{n, n - \tau},
\end{equation}
which follows from Bohr's postulates. By the correspondence principle, since a quantum theoretic frequency is related to the classical frequency as
\begin{equation}
    \nu(n, \tau) \leftrightarrow \nu_{n, n - \tau}.
\end{equation}
At that point, he argued that the corresponding amplitude in the quantum case should be
\begin{equation}
    x(n, \tau) \leftrightarrow x_{n, n - \tau}.
\end{equation}
Then the set $\{ x_{n, n - \tau} \exp{[2\pi i \nu_{n, n-\tau}t]} \}$ may be chosen as the representative of $x_n(t)$ in the quantum case. Now, in the classical case
\begin{equation}
    \begin{split}
        x_n(t)^2 & = \left[ \sum_\tau x(n, \tau) \exp{[ 2\pi i \nu(n, \tau) t ] } \right] \left[ \sum_{\tau'} x(n, \tau') \exp{[ 2\pi i \nu(n, \tau') t ] } \right] \\
        & = \sum_{\tau \tau'} x(n, \tau) x(n, \tau')\exp{[ 2\pi i \nu(n, \tau + \tau')t ]}
    \end{split}
\end{equation}
by \eqref{classical frequency relation}. By changing variables as $\tau \to \tau - \tau'$, we get
\begin{equation}
    \sum_{\tau = -\infty}^{+\infty} x^2(n, \tau) \exp{[ 2\pi i \nu(n, \tau) t ] },
\end{equation}
where
\begin{equation}
    x^2(n, \tau) = \sum_{\tau' = -\infty}^{+\infty} x(n, \tau')x(n, \tau - \tau').
\end{equation}
Therefore, according to \eqref{quantum mechanical frequency sum rule}, the corresponding quantum quantity must be
\begin{equation}
    x^2_n(t) \leftrightarrow \{ x^2_{n, n - \tau } \exp{[ 2\pi i \nu_{n, n -\tau} t ]} \}.
\end{equation}
To get the correct phase factor, one must replace the terms like $n, n-\tau' \times n, n -(\tau - \tau')$ by $n, n - \ tau' \times n - \tau', n - \tau$. Accordingly, we have
\begin{equation}
    x^2_{n, n - \tau} = \sum_{\tau'} x_{n, n - \tau'} x_{n - \tau', n - \tau}.
\end{equation}
The generalization of this rule as $x_n(t) \eta_n$ yields a noncommutative algebra, since, in general
\begin{equation}
       \sum_{\tau'} x_{n, n - \tau'} y
       _{n - \tau', n - \tau} \neq \sum_{\tau'} y_{n, n - \tau'} x_{n - \tau', n - \tau}.
\end{equation}
As it was said in earlier chapters, in the old theory, to solve the quantum problems, one had had to first calculate the equation of motion classically and then impose the quantum of action condition
\begin{equation}
\label{actionprinciple}
    \oint p dq = nh,
\end{equation}
which worked for some basic problems but was hard to work out for most others. So, for a generic equation of motion $\ddot{x} + f(x) = 0$, $p = m \dot{x}$, and by the correspondence principle
\begin{equation}
    \tau \frac{\partial \Phi(n, \tau)}{\partial n} \leftrightarrow \Phi_{n + \tau, n} - \Phi_{n, n - \tau},
\end{equation}
Heisenberg showed that
\begin{equation}
    \begin{split}
        \oint p dq & = \int_0^{1/\nu} m \dot{x}^2 dt = 2 \pi^2 m \sum_\tau \tau \nu(n, \tau) |x(n, \tau)|^2 = nh \\
        & \Longrightarrow h = 2 \pi^2 m \sum_\tau \tau \frac{d}{dn}(\nu(n, \tau) |x(n, \tau)|^2) = h\\
        & \leftrightarrow  2 \pi^2 m \sum_\tau \nu_{n+\tau, n} |x_{n + \tau, n}|^2 - \nu_{n, n - \tau} |x_{n, n - \tau}|^2 = h,\\
    \end{split}
\end{equation}
where $x(n, \tau) = x^*(n, -\tau)$ since $x_n(t)$ has to be real,\footnote{Consider the elements corresponding to $\tau = a$ and $\tau = -a$ in the sum,
\begin{equation}
    \begin{split}
        & x(n, a)\exp{[2\pi i\nu(n, a)t]} + x(n, -a)\exp{[2\pi i\nu(n, -a)t]} \\
        = & x(n, a)\exp{[2\pi i\nu(n, a)t]} + \{x(n, a)\exp{[2\pi i\nu(n, a)t]\}^* },
    \end{split}
\end{equation}
which real.
} 
is the quantum condition for such a system. Therefore, by solving the equation of motion and applying the quantisation rule that he derived, he argued that the frequencies and the transition amplitudes, which are observable, can be calculated from this mathematical scheme. Afterwards, he went on to apply the scheme to the quantum simple anharmonic oscillator.\footnote{The details of the derivation are not clear in Heisenberg's paper, \cite{Heisenberg1925}. For a clearer account, see \cite{Aitchison2004}. }

\subsection{ Born and Jordan's Contributions }
It was Jordan and Born with whom Heisenberg improved the theory into matrix mechanics \cite{BornHeisenbergJordan1926, BornJordan1925}. The two profound papers that they published included many results, taking the theory into its full-fledged matrix mechanics form. Starting with the quantum of action principle and by the Fourier expansions of $p$ and $q$, through a similar calculation to which was just presented, they got the canonical commutation relation.

Let 
\begin{equation}
    p = (p(nm)e^{2\pi i v(nm)t}), \: q = (q(nm)e^{2\pi i \nu(nm)t}),
\end{equation}
which are Hermitian. We also assume that the following relations hold
\begin{equation}
    \begin{split}
        & q(nm)q(mn) = |q(nm)|^2, \\
        & \nu(nm) = \nu(mn), \\
        & \nu(jk) + \nu(kl) + \nu(lj) = 0.
    \end{split}
\end{equation}
By using the second and third relations, we get
\begin{equation}
    \begin{split}
        & \nu(jk) = - \nu(kl) - \nu(lj) = \nu(jl) - \nu(kl) \\
        \Longrightarrow & h\nu(jk) = h(\nu(jl) - \nu(kl)) = W_j - W_k.
    \end{split}
\end{equation}
Consider a matrix of the form $g(pq) = (g(nm)e^{2\pi i \nu(nm)t}) \equiv (g(nm))$. The time derivative yields
\begin{equation}
    \dot{g} = 2\pi i (\nu(nm)g(nm)).
\end{equation}
Since $\nu(nm) \neq 0$ for $ n \neq m$, if $\dot{g} = 0$, then $g$ is necessarily diagonal.

Now, using \eqref{actionprinciple} and \eqref{correspondance principle}, along with the fact that
\begin{equation}
    d = pq - qp \Rightarrow \dot{d} = 0,
\end{equation}
(by another matrix theory-related result in the paper), so that the matrix $d$ is diagonal, we have the canonical commutation relation:
\begin{equation}
\label{canonical commutation relation}
    pq - qp = -\frac{i h}{4 \pi} I.
\end{equation}
Subsequently, they also derive that
\begin{equation}
\label{heisenberg equation}
    \dot{g} = -\frac{2 \pi i}{h}[g, H],
\end{equation}
and thus, for $g = H$, that the energy is conserved (for $\partial_t H = 0$ of course), which also entails the result that $H$ is diagonal on the base which it shares with $p, q$.

Soon after, Wiener and Born introduced the operator algebra into quantum theory \cite{BornWiener1926}. For $y(t) = \sum_m y_m e^{2 \pi i W_m t/h}$ and $x(t) = \sum_m x_n(t) e^{2 \pi i W_n t/h}$, we have
\begin{equation}
    x_n(t) = \lim_{T \to \infty} \frac{1}{2T} \int_{-T}^{T} x(t)e^{-2\pi i W_n t/h}dt,
\end{equation}
and, wherefore, the transformation $y_m = \sum_n q_{mn} x_n(t)$ entails the equation
\begin{equation}
    y(t) = \lim_{T \to \infty} \frac{1}{2T} \int_{-T}^{T} \left[ \sum_{mn} q_{mn} e^{2\pi i (W_m t - W_n s)/h} \right] x(s) ds.
\end{equation}
Writing 
\begin{equation}
\label{q}
    q = \lim_{T \to \infty} \frac{1}{2T} \int_{-T}^{T} \underbrace{\left[ \sum_{mn} q_{mn} e^{2\pi i (W_m t - W_n s)/h} \right]}_{q(s,t)} ds,
\end{equation}
the former expression can be put into the form
\begin{equation}
    y(t) = qx(t).
\end{equation}
As an example, direct calculation with the linear operator $D = d/dt$, yields
\begin{equation}
    y(t) = Dqx(t) \Rightarrow \frac{\partial q(t,s)}{\partial t} = \frac{2\pi i}{h} \sum_{mn}q_{mn} W_m e^{2\pi i (W_m t - W_n s)/h}.
\end{equation} 
Therefore
\begin{equation}
\begin{split}
    & (Dq)_{mn} = \frac{2\pi i}{h} q_{mn} W_m, \: \underbrace{(qD)_{mn} = \frac{2\pi i}{h} q_{mn} W_n}_{\text{By partial integration}} \\
    &\Longrightarrow [D,q]x = \frac{2 \pi i}{h}(W_m - W_n)q_{mn}x = 2\pi i \nu_{mn}q_{mn}x. \\
\end{split}
\end{equation}
Hence, by definition of $q$ in \eqref{q}, the equation just derived entails
\begin{equation}
    \Longrightarrow [D, q] = \dot{q}.\footnotemark
\end{equation}
By \eqref{heisenberg equation}, we also have the familiar result $D = d/dt = \frac{2 \pi i}{h}H$.
\footnotetext{In operator algebra, for an arbitrary function $x$, this equation can also be derived as
\begin{equation}
    [D, q]x = \frac{d(qx)}{dt} - q\frac{dx}{dt} = \frac{dq}{dt}x + q\frac{dx}{dt} - q\frac{dx}{dt} = \dot{q}x.
\end{equation}}

\subsection{The Correspondance Between the Classical and Quantum}
At about the same time, the Göttingen school put the matrix mechanics forward, Dirac found a mathematical equivalence between classical and quantum mechanics \cite{Dirac1925}. Dirac started by defining $d/dv$ as the "most general operator". By assumption, it obeys the following rules:
\begin{enumerate}[label = \Roman*.]
    \item \begin{equation}
        \frac{d}{dv}(x + y) = \frac{dx}{dv} + \frac{dy}{dv},
    \end{equation}
    \item \begin{equation}
        \frac{d(xy)}{dv} = x\frac{dy}{dv} + \frac{dx}{dv}y,
    \end{equation}
\end{enumerate}
where $x, y$ obey the matrix product rule, $(xy)_{mn} = \sum_k x_{mk}y_{kn} $. Then, by the first rule, 
\begin{equation}
\label{Dx_mn}
    (Dx)_{mn} = (\frac{dx}{dt})_{mn} \equiv \sum_{m'n'} a_{nm, n'm'} x_{n'm'}.
\end{equation}
Therefore, the second rule can be written as
\begin{equation}
    \sum_{n'm'k} a_{nm, n'm'} x_{n'k} y_{k m'} = \sum_{kn'k'} a_{nk, n'k'} x_{n'k} y_{k m'} + \sum_{kk'm'} x_{nk} a_{km, k'm'} y_{k'm'}.
\end{equation}
Comparing the amplitudes, Dirac reduced \eqref{Dx_mn} to
\begin{equation}
    (\frac{dx}{dv})_{mn} = \sum_k x_{nk} a_{km} - a_{nk} x_{km}
\end{equation}
or
\begin{equation}
    \frac{dx}{dv} = [x,a],
\end{equation}
which showed that the most general operation on a "quantum variable" $x$, subject to the rules presented at the beginning, is performed by "taking the difference of its Heisenberg products with some other quantum variable" (Dirac, 1925, p.647).

To see what this equation corresponds to in classical mechanics, Dirac made use of the correspondence principle \eqref{correspondance principle}, which resulted in
\begin{equation}
\label{correspondance between commutation relations}
    [x, y] \leftrightarrow \frac{i h}{2 \pi} \{ x, y \},
\end{equation}
where
\begin{equation}
    \{ x, y \} = \frac{\partial x}{\partial q} \frac{\partial y}{\partial p} - \frac{\partial y}{\partial q}\frac{\partial x}{\partial q}
\end{equation}
is the Poisson bracket.\footnote{Suppose $f(p,q, t)$ is a solution to a system defined by the Hamiltonian $H(p, q, t)$. The chain rule reads
\begin{equation}
    \frac{df}{dt} = \frac{\partial f}{\partial q} \frac{dq}{dt} + \frac{\partial f}{\partial p} \frac{dp}{dt} + \frac{\partial f}{\partial t}.
\end{equation}
If the system is described by $p$ and $q$ s.t. they obey Hamilton's equations
\begin{equation}
    \begin{split}
        & \dot{q} = \frac{\partial H}{\partial p}, \\
        & \dot{p} = -\frac{\partial H}{\partial q},
    \end{split}
\end{equation}
we have
\begin{equation}
    \dot{f} = \{ f, H \} +\partial_t H.
\end{equation}
From this, for an operator $Q$, using Dirac's correspondence principle, we get the Heisenberg equation of motion
\begin{equation}
    \dot{Q} = \frac{2 \pi i}{h}[Q, H] + \partial_tH.
\end{equation}

For more information on canonical transformations and Poisson brackets, see the second appendix and \cite{Goldstein2002Classical}.}
In classical mechanics, for a system described by the set of coordinates $\{ q_i, p_i \}$, the canonical commutation relations are
\begin{equation}
    \{ q_k, p_l \}_{qp} = \sum_i \frac{\partial q_k}{\partial q_i} \frac{\partial p_l}{\partial p_i} - \frac{\partial q_k}{\partial p_i} \frac{\partial p_l}{\partial q_i} = \delta_{kl},
\end{equation}
where $\{ p_k, p_l \} = \{ p_k, p_l \} = 0$. Such coordinates are called canonical coordinates, and the transformations that transform one canonical set of coordinates to another (i.e., the coordinates which obey the canonical commutation rules) are called the canonical transformations. The Poisson brackets of canonical coordinates are invariant under these transformations, and so are the equations of motion. The problems in quantum mechanics can be thus reduced to finding suitable canonical coordinates (ones with the most constants of motion) and then transforming the equations of motion using the Dirac correspondence principle, as is shown in \cite{BornHeisenbergJordan1926}. It is interesting to see that the canonical transformations correspond to unitary transformations in quantum mechanics.

In conclusion, the main mathematical connection between classical and quantum mechanical equations, in the case of matrix mechanics, was the correspondence principle. It was included in the main formulation of the theory by Heisenberg and made an implicit feature by the contributions of Born and Jordan in matrix mechanics in the equations \eqref{canonical commutation relation} and \eqref{heisenberg equation}. The same principle appeared in the correspondence between the commutation relations and Poisson brackets, as shown in \eqref{correspondance between commutation relations}. At that point, we can as well say that one of the foundational postulates of quantum theory was this principle. However, in wave mechanics, we will not see the principle in direct use. Instead, the main postulate of the theory will be that particles have some kind of an "intrinsic periodic phenomena" which causes them to act like waves. 

\section{Wave Mechanics}
\subsection{Preliminary Considerations}
A significant work on the reconciliation of wave and particle phenomena was from Hamilton, whose main concern while studying mechanics and deriving his equations was to find a general method to explain optical phenomena, as Lagrange did for mechanics \cite{Hamilton1833}.\footnote{From \cite{Hamilton1833}, p.5: "Those who have meditated on the beauty and utility, in theoretical mechanics, of the general method of Lagrange—who have felt the power and dignity of that central dynamical theorem which he deduced, in the M'echanique Analytique, from a combination of the principle of virtual velocities with the principle of D'Alembert—and who have appreciated the
simplicity and harmony which he introduced into the research of the planetary perturbations, by the idea of the variation of parameters, and the differentials of the disturbing function, must feel that mathematical optics can only then attain a coordinate rank with mathematical mechanics, or with dynamical astronomy, in beauty, power, and harmony, when it shall possess an appropriate method, and become the unfolding of a central idea."}
Yet his main interest was in mathematical laws governing both phenomena rather than a direct statement about the corpuscular nature of light or wave nature of matter \cite{JoasLehner2007}.

The analogy in mathematical form can be summarised as follows: The principle equation in ray optics is Fermat's principle of least time,
\begin{equation}
    \delta T = \delta \int \frac{n}{c} ds = 0,
\end{equation}
whereas in mechanics, it is the Maupertuis' principle of least action,
\begin{equation}
    \delta S = \delta \int \sqrt{2m(E-U)}ds,
\end{equation}
which are just special cases of Hamilton's principle of stationary action,
\begin{equation}
    \delta S = \delta\int L dt = 0,
\end{equation}
where $L$ is the Lagrangian. This analogy can be grounded by observing that the wave equation converges to the Eikonal equation for large wavelengths, which is the fundamental equation of geometrical optics. The calculation is as follows: For $\psi = A e^{ik_0 W}$, where $A$ and $W$ are slowly varying functions of position, the wave equation $\Delta \psi + n^2 k_0^2 u = 0$ reads
\begin{equation}
    \Delta (Ae^{ik_0W}) + n^2k_0^2Ae^{ik_0W} = 0.
\end{equation}
For simplicity, let's take $A$ and $W$ as functions of only $x$. Therefore,
\begin{equation}
            \frac{\partial }{\partial x}\left[ \partial_x A e^{ik_0W} + Aik_0\partial_x W e^{ik_0W} \right] + n^2k_0^2Ae^{ik_0W} = 0
\end{equation}
\begin{equation}
    \begin{split}
        \Longrightarrow \partial_x^2A + \partial_x A & \partial_x W ik_0 + \partial_xA\partial_x W ik_0 \\
        &+ A \partial_x^2W ik_0 - A(\partial_xW)^2 k_0^2 = -n^2k_0^2A
    \end{split}
\end{equation}
\begin{equation}
    \Longrightarrow (\partial_x W)^2 = n^2,
\end{equation}
where we neglected the lower powers of $k_0$ by the assumption we made. Therefore, as geometrical optics is the long wavelength approximation of undulatory optics, maybe the particle phenomena were just a similar approximation of the wave phenomena.

The phenomena of radiation as a wave or particle were still topics of discussion. Some phenomena, like black body radiation and the photoelectric effect, were explained by corpuscular theory, whereas others, like refraction, diffraction, etc., were explained by wave mechanics. It was Louis de Broglie who first introduced the idea of associating waves with particles in the context of quantum theory \cite{deBroglie1924}. It seems like he wasn't aware of Hamilton's optical-mechanical analogy, but he was deeply impressed by the works of Marcel Brillouin (Jammer 1989, p.246-247).

\subsection{De Broglie's Idea of Matter-Waves}
Consider a particle of rest-mass $m$ with an intrinsic periodic motion. It has a frequency $\nu_0 = mc^2/h$ at rest. Relative to an observer, for which the particle moves with the velocity $v = \beta c$, the frequency is
\begin{equation}
    \nu = \frac{mc^2}{h\sqrt{1 - \beta^2}} = \frac{\nu_0}{\sqrt{1- \beta^2}}.
\end{equation}
according to its energy content. But one also expects the frequency of the internal phenomenon to change according to time dilation as
\begin{equation}
    \nu_1 = \nu_0\sqrt{1-\beta^2},
\end{equation}
so, there should be a relation such that
\begin{equation}
    \nu_1 = \nu(1-\beta^2).
\end{equation}
The apparent discrepancy is lifted when one considers the following: Let there be a wave with frequency $\nu$, which spreads with the velocity $c/\beta = c^2/v$, and a particle with the intrinsic periodic phenomena coincided at $t = 0$, to which we just ascribed the frequency $\nu_1$.\footnote{The reason why frequencies are ascribed such that can be seen by the fact that the first one satisfies the Doppler effect.} The wave is described by the equation $\sin{2\pi \nu x (1/v - \beta /c)}$ after a time $t$, and the internal motion changes with respect to the equation $\sin 2\pi \nu_1 x/v$. Therefore,
\begin{equation}
    \nu x(1/v - \beta /c) \stackrel{!}{=} \nu_1 x/v \Rightarrow \nu_1 = \nu(1-\beta^2),
\end{equation}
as excepted. Therefore, "any moving body may be accompanied by a wave and that it is impossible to disjoin motion of body and propagation of wave" (de Broglie, 1924, p.450). The velocity $c/\beta$ is the phase velocity of the wave. The phase velocity and group velocity are related by
\begin{equation}
    1/v_g = d(\nu/v)/d\nu = 1/\beta c) = 1/v,
\end{equation}
so $v_g = v$, that is, the group velocity of the wave is the velocity of the particle.

Therefore, de Broglie suggested that the wave phenomena is present when the means of measurement is on the scale $\lambda$ of the matter waves, whereas otherwise the structure of the wave goes undisturbed and what we measure is related to corpuscular phenomena. He also suggested that the probability of absorption or scattering of light quanta is related to the phase wave vectors crossing upon the atom.

One of the results that de Broglie proposed was that the Sommerfeld quantum condition $\oint p dq = nh$, along with $\lambda = h/p$ is equivalent to
\begin{equation}
    \oint \lambda^{-1}dq = n,
\end{equation}
which states that the (monochromatic) wavelength of the electron in an orbit is quantised as $2\pi r/n$, resembling the solution of a boundary value problem, attracted the interest of Schrödinger deeply. That was also the wave-mechanical explanation of Bohr's quantum condition of discrete angular momentum of the electron in the orbits.

\subsection{The Wave Equation}
Thinking that if there are waves, there should also be a wave equation, Schrödinger began his study on de Broglie's matter waves for bound particles. He thought that for resolutions much higher than the associated wavelength $\lambda = h/p$, just as in geometrical optics, the wavelike phenomena were not possible to observe, whereas, in the high resolutions, i.e. where the interaction distances $d$ between the particles were $d \approx \lambda$, the wavelike nature of the matter was significant to the behaviour of the system, just as in undulatory optics. So, just as de Broglie thought, he showed that there was more to the resemblance between the analogy with geometrical-undulatory optics and that of the particle-wave nature of the matter \cite{Schrodinger1926Undulatory}.

One of the works that impressed Schrödinger the most was the papers of Einstein on the ideal gas, where he used Bose's counting procedure \cite{Bose1924}, which renders the particles indistinguishable and thus arranges them in the possible energy states according to the Ehrenfest and Krammling's combinatoric expression, to arrive at Planck's formula without the use of classical electrodynamics, contrary to the way Planck derived it \cite{Einstein_1924, einstein1925}. Schrödinger saw that the wave concept presented a natural way to account for the phenomenon \cite{JoasLehner2007}.

Schrödinger published four papers (and a paper to summarise the results) where he studied the form of the wave equation and applied it successfully to the problems of quantum mechanics \cite{Schrodinger1926Undulatory, Schrodinger1926Eigenwert1, Schrodinger1926Eigenwert2, Schrodinger1926Eigenwert3, Schrodinger1926Eigenwert4}.\footnote{His works assumes some knowledge on Hamilton-Jacobi formulation of classical mechanics, a summary of which can be found in the second appendix.} He started his first communication \cite{Schrodinger1926Eigenwert1} by proposing that in the time-independent Hamilton equation (since he was dealing with stationary states in the Hydrogen atom),
\begin{equation}
    H(q, \frac{\partial W}{\partial q}) = E.
\end{equation}
Hamilton's characteristic function $W$ is completely separable for orthogonal coordinates and quadratic (in the generalised momenta) Hamiltonians with no time dependence if the potential energy is additively separable in each coordinate.\footnote{See the second appendix, section B.1.} Therefore, he assumed
\begin{equation}
\label{ansatz S = Klnpsi}
    W = \sum_i W_i(q_i;\partial_{q_1}W,\dots,\partial_{q_n}W) \equiv K\ln\left( \prod_j \psi (q_j) \right) = K\ln \psi,
\end{equation}
\begin{equation}
\label{gleichung 1'}
    \Longrightarrow H\left(q, \frac{K}{\psi}\frac{\partial \psi}{\partial q} \right) = E.
\end{equation}
where $K$ is a constant, and $\psi$ are single-valued functions of only one of the coordinates $q_j$, which are real (this condition is lifted in the subsequent papers). In simple cases, this equation can be written in quadratic form equated to zero. In this case, $\psi$ also has to be twice continuously differentiable, for which the first derivative is equal to zero for the equation to be quadratic, which extremizes the corresponding integral equation.\footnote{The last condition is rather arbitrary, as Schrödinger will also give in in his second communication, but surprisingly leads to the correct answer, the reason of which will be given in the subsequent pages. The motivation behind this rather \textit{ad hoc} principle was argued to be related to optical-mechanical analogy (Joas, Lehner, 2007, p.13):
\begin{quote}
    After several unsuccessful attempts at guessing the variational principle, there appears in the notebook a program that
contains as its second item "the old Hamiltonian analogy between optics and mechanics". In
the text corresponding to this item, Schrodinger starts from the Hamiltonian partial differential
equation and reinterprets it as a variational principle which indeed leads him to the (nonrelativistic) wave equation.
\end{quote}
} Therefore, he effectively changed the quantum conditions with the conditions on $\psi$.\footnote{From \cite{Schrodinger1926Eigenwert1}, p.1: "Wir suchen nun nicht eine Lösung der Gleichung \eqref{gleichung 1'}, sondern wir stellen folgende Forderung. Gleichung \eqref{gleichung 1'} lässt sich bei Vernachlässigung der Massenveränderlichkeit stets, bei Berücksichtigung derselben wenigstens dann, wenn es sich um das Einelektronenproblem handelt, auf die Gestalt bringen: quadratische Form von $\psi$ und seinen ersten Ableitungen $= 0$. Wir suchen solche reelle, im ganzen Konfigurationsraum eindeutige, endliche und zweimal stetig differenzierbare Funktionen $\psi$, welche das über den ganzen Konfigurationsraum erstreckte Integral der eben genannten quadratischen Form zu einem Extremum machen. Durch dieses Variationsproblem ersetzen wir die Quantenbedingungen."}

In cartesian coordinates, where $e,m$ are the electron's mass and charge, according to our considerations in the second appendix, the Hamilton-Jacobi equation takes the form
\begin{equation}
    \sum_{i=1}^3 \frac{p^2_i}{2m} + V(x,y,z) = \frac{K^2}{2m} \sum_{i=1}^3 \left( \frac{1}{\psi} \frac{\partial\psi}{\partial q_i} \right)^2 - \frac{e^2}{r}  = E
\end{equation}
\begin{equation}
    \Longrightarrow \sum_{i=1}^3 \left( \frac{\partial\psi}{\partial q_i} \right)^2 - \frac{2m}{K^2}\left( E + \frac{e^2}{r} \right)\psi^2  = 0,
\end{equation}
where $r = \sqrt{x^2 + y^2 + z^2}$, and $\frac{K}{\psi}\frac{\partial \psi}{\partial q_i} = p_i$ by \eqref{p_i = partial F_2}. Therefore, our variation problem becomes
\begin{equation}
    \delta J = \delta \int^{+\infty}_{-\infty} d^3q \sum_{i=1}^3 \left( \frac{\partial\psi}{\partial q_i} \right)^2 - \frac{2m}{K^2}\left( E + \frac{e^2}{r} \right)\psi^2 = 0,
\end{equation}
from which we get the following equations,
\begin{equation}
    \begin{split}
        & \Delta \psi + \frac{2m}{K^2} \left( E + \frac{e^2}{r}\right)\psi = 0  ,\\
        & \int df \delta \psi \frac{\partial \psi}{\partial n} = 0,
    \end{split}
\end{equation}
of which the first is the Schrödinger equation for an electron in the hydrogen atom, and the second, where $df$ is the surface element of a surface of infinite extent, is the condition that the wavefunction must go to zero at infinity.\footnote{In reference \cite{ip2013mystery}, there is an account which explains how this variational principle works by means of statistical considerations. The number of states in an imaginary ensemble in the configuration space, which is encoded by the "field" $W$, is
\begin{equation}
    N = \int \rho(q_i,)dq_1 \dots dq_n,
\end{equation}
for $n$ degrees of freedom, and $\rho(q_i, t)$ is the density of states in the ensemble. Now, if we apply the normalisation condition, $N = 1$, then we have a probability distribution, and we can define the probability of finding the system at the state $q_i$ as
\begin{equation}
    P(q_i, t) = \frac{\rho(q_i,t)}{\int \rho(q_i,)dq_1 \dots dq_n} = \rho(q_i,t).
\end{equation}
The evolution of the probability distribution is given by the continuity equation,
\begin{equation}
    \frac{\partial \rho}{\partial t} + \nabla \cdot \left( \rho \frac{\nabla W}{m} \right) = 0.
\end{equation}
Later in the paper, they state that the statistical Hamilton-Jacoby theory, as they call it, can be summarised by the following variational principle,
\begin{equation}
    \delta 
\int dq_1 \dots dq_n \,
\rho \left[ \frac{\partial W}{\partial t} + H \left( q_i, \frac{\partial W}{\partial q_i} \right) \right] = 0,
\end{equation}
by treating $\rho$ and $W$ as conjugate variables. The Hamilton-Jacobi equation and the continuity equation are then derived from this variational principle. It can be seen, upon comparison, that this variational principle is in the same form as Schrödinger's variational principle, for $\rho = |\psi|^2$. By the ansatz $W = K \ln \psi$, we have
\begin{equation}
    \psi = \sqrt{\rho} e^{W/K},
\end{equation}
where $K$ has to be an imaginary number to satisfy $\rho = |\psi|^2$.
}

Therefore, Schrödinger reduced to problem of finding the energy levels of the electron in the hydrogen atom to a boundary value problem for negative $E$, i.e., bound states. For positive $E$, which classically corresponds to hyperbolic solutions of the Kepler problem, there are no bound states as the electron is not bounded by the potential. For bound states, he showed that quantisation naturally occurred whenever
\begin{equation}
    n = \frac{me^2}{K\sqrt{-2mE}}
\end{equation}
is a real positive integer, to give energy eigenvalues of
\begin{equation}
    E = -\frac{me^4}{2K^2n^2},
\end{equation}
which provides the Bohr energy levels for $K = h/2\pi$.

At the beginning of the second communication \cite{Schrodinger1926Eigenwert2}, Schrödinger gives another means of derivation for the wave equation, as the ansatz \eqref{ansatz S = Klnpsi} and the variational principle he postulated were "unintelligible" (unverständlich). This time, he explicitly uses the optical-mechanical analogy:
\begin{quote}
    Maybe our classical mechanics is the full analogue of geometrical optics, and, as such,
wrong, not in agreement with reality. It fails as soon as the radii of curvature and the
dimensions of the trajectory are not large anymore compared to a certain wavelength,
to which one can attribute a certain reality in q-space. In that case, one has to
search for an "undulatory mechanics"—and the obvious way to this end is the wave theoretical extension of Hamilton's picture. (Schrodinger, \cite{Schrodinger1926Eigenwert2}, pp. 25)\footnote{Translation is from \cite{JoasLehner2007}, p.15.}
\end{quote}
For a conservative system, the action is
\begin{equation}
    S = \int T - U dt,
\end{equation}
 and we have the equation
\begin{equation}
\label{hje}
    \frac{\partial S}{\partial t} + \frac{1}{2m}|\nabla S|^2 + U = 0.
\end{equation}
Substituting the usual ansatz
\begin{equation}
    S(q_i,t) = -Et + W(q_i),
\end{equation}
in the equation \eqref{hje}, and by using the fact that $|\nabla S| = |\vec{P}|=P$, we get
\begin{equation}
    |\nabla S| = \sqrt{2m(E - U)}, \: \partial_t S = -E,
\end{equation}
from which follows, for constant action,
\begin{equation}
    \begin{split}
        dS = &|\nabla S|ds  + \partial_t S dt = (V |\nabla S| - E)dt = 0 \\
        & \Longrightarrow V = \frac{E}{|\nabla S|} = \frac{E}{\sqrt{2m(E - U)}}, \\
    \end{split}
\end{equation}
where $V$ is the phase velocity of the waves. On the other hand, the velocity of the particle is 
\begin{equation}
    |\nabla S|/m = P/m = v = \sqrt{2(E-U)/m}.
\end{equation}
Therefore, he showed the equivalence of Fermat's and Maupertuis' principle:
\begin{equation}
    \begin{split}
        &     \delta \int ds/V = \delta \int ds \: \sqrt{2m(E - U)}/E = \delta \int vdt \sqrt{2m(E - U)}/E \\
        & = \delta \int 2(E - U)/E dt = 0 \Longrightarrow \delta \int 2T dt  = 0. \\
    \end{split}
\end{equation}
Afterwards, he introduced the waves
\begin{equation}
    \psi = A(x,y,z) \sin(W/K),
\end{equation}
where $K$ is a constant with the units of action. In this case, the frequency is
\begin{equation}
    \nu = \frac{E}{2\pi K},
\end{equation}
which yields the relation of Einstein for $K = h/2\pi$. The associated wavelength is given as
\begin{equation}
    \lambda = v/\nu = h/\sqrt{2m(E-U)} = h/p.
\end{equation}
(Note that the equality of particle velocity and the group velocity of the wave does also hold, as we have seen in de Broglie's papers.) Next, he introduced the ordinary wave equation
\begin{equation}
    \Delta \Psi - \frac{\ddot{\Psi}}{V^2} = 0,
\end{equation}
with the condition that
\begin{equation}
    \Psi(q,t) = \psi(q)e^{2 \pi iEt/h},
\end{equation}
he got the wave equation\footnote{In \textit{Atombau und Spektrallinien}, \cite{sommerfeld1921}, Sommerfeld used the optical-mechanical analogy to derive Schrödinger's equation as follows. Starting with the equation
\begin{equation}
    \frac{1}{2m}(p_x^2 + p_y^2 + p_z^2) = E - V = \frac{|\nabla S|^2}{2m},
\end{equation}
where $S$ is Hamilton's principle function, and compared it with the wave equation
\begin{equation}
    \Delta u = \frac{1}{v^2}\frac{\partial^2 u}{\partial t^2}.
\end{equation}
Next, we use the ansatz $u = \psi e^{i\omega t}$, where $\omega = k/v $, and $ k = nk_0$, where $k_0$ is the wavelength in the vacuum and $n = c/v$ is the refractive index. Therefore we get
\begin{equation}
    \Delta \psi + n^2k_0^2\psi = 0.
\end{equation}
Next, he shows the approximate relation between this equation and the Eikonal equation, as it is done in preceding pages (instead of $W$, he uses $S$). Therefore, by the relations
\begin{equation}
    |\Delta S|^2 = 2m(E-V), \: |\Delta S|^2 = n^2,
\end{equation}
we get
\begin{equation}
    \Delta \psi + 2m(E - V)k_0^2 \psi = 0.
\end{equation}
But in the first equation, $S$ is dimensionless, whereas in the second one, it has dimensions of energy and mass. The equality of dimensions is retrieved once we state $k_0 = (2\pi/h)^2$.
}:
\begin{equation}
    \Delta \psi + \frac{8 \pi^2 m}{h^2}(E - U)\psi = 0.
\end{equation}
This derivation of the wave equation is what he also preferred in \cite{Schrodinger1926Undulatory}.

In the third communication \cite{Schrodinger1926Eigenwert3}, of which we will not delve into the details, he developed the time-independent perturbation theory and applied it to calculate the Stark effect of the hydrogen atom.

In the last paper of the series \cite{Schrodinger1926Eigenwert4}, the fourth communication, he developed the time-dependent perturbation theory for non-conservative systems. Of course, to do that, he needed the time-dependent wave equation, as the wave equation derived and used until now was for conservative systems, especially for the case of the stationary states of the electron in the hydrogen atom. For time-dependent potentials, the theory had to be extended. To do that, Schrödinger observed that
\begin{equation}
     -\frac{ih}{2\pi} \partial_t \Psi = E \Psi, 
\end{equation}
and postulated that the equation
\begin{equation}
    -\frac{h^2}{8\pi^2m}\nabla^2\Psi + U\Psi = \frac{-ih}{2\pi}\frac{\partial \Psi}{\partial t}.
\end{equation}
For that matter, he also lifted the reality condition on $\psi$.\footnote{As expected, it is possible to derive Hamilton-Jacobi equation as a classical approximation of this equation, by letting $\psi = \sqrt{\rho}\exp iS(x, t)/\hbar$, where $\hbar = h/2\pi$, substituting that into the Schrödinger equation, and then taking the limit $\hbar \to 0$, which gives
\begin{equation}
    \frac{1}{2m} \left( \frac{\partial S}{\partial x} \right)^2 + V(x) + \frac{\partial S}{\partial t} = 0,
\end{equation}
in one dimension.
}

Although Schrödinger believed in the literal existence of matter waves, the wave function that he got was not suitable for such tangible existence, for it was not defined in 3-dimensions. For a system of $N$ particles, it is defined on the $3N$-dimensional configuration space.

Schrödinger's idea of the primacy of waves over particles did not become a consensus. In 1926, just after Schrödinger published his papers on wave mechanics, Born suggested his statistical interpretation of the wave function in his studies on the scattering theory \cite{born1926b, born1926a}. Schrödinger strongly resisted Born's interpretation, believing that wave mechanics should describe real physical waves rather than probabilities. However, the success of the probabilistic interpretation in scattering experiments ultimately led to its acceptance and continued to defend his position throughout his life.

\section{Uncertainty Principle, Indistinguishability, and Complementarity}
Although the formal language of quantum theory was slowly gaining ground, the philosophical implications were still hot topics of discussion. Heisenberg, Born and others besides these were favouring the particle view over the wave concept, accepting the quantum jumps as they are,  whereas Schrödinger, Einstein and other physicists were repulsed by these ideas. Observing that the problem lies in transforming classical concepts like position and velocity into the language of quantum physics, in a paper in 1927, Heisenberg set out to find a way to describe these concepts operationally, as Einstein did to space and time \cite{heisenberg1983}.

For Heisenberg, for an observable to have a meaningful description, the measurement process and result have to be defined clearly. For instance, the concepts of space and time according to Newton do not have a clear means of measurement, whereas Einstein's are defined directly on measurements. For that matter, he considered an experiment with $\gamma$-rays to determine the position of an electron. The velocity of the electron will change according to the wavelength of the light, which in turn affects the accuracy of the measurement. The photon's wavelength will change according to the Compton effect as
\begin{equation}
    \Delta \lambda = \frac{h}{m_e c}(1 - \cos \theta).
\end{equation}
However, since the scattering angle is not known with certainty, we cannot know the wavelength shift and, therefore, the momentum of the scattered photon. After these considerations, he derived the result:
\begin{equation}
    \Delta p \Delta q \gtrsim h,\footnotemark
\end{equation}
\footnotetext{He derived the exact expression 
\begin{equation}
    \Delta p \Delta q \geq \hbar/2,
\end{equation}
for only Gaussian states. The formula is later shown to hold in general by Kennard \cite{kennard1927}.
}This thought experiment seems to be valid for classical particles, too. However, in classical picture, the particles have exact positions and momenta before measurement, whereas in quantum theory, they do not.\footnote{In wave picture, this principle arises from the inverse behaviour of spreading of wave functions related to conjugate variables.} So, it is not a particularly good example of explaining the inherent uncertainty in nature. Generalisations of the principle for any two conjugate variables followed by others.\footnote{A counter-argument came from Condon, \cite{Jammer1966}, p. 353 - 354, in the meantime. The basic idea can be outlined as follows: Consider the angular momentum of the electron in the hydrogen atom, $L_z = m \hbar$, which entails $\Delta L_z = 0$, and therefore $\Delta L_z \Delta L_x = 0$, although $ [L_z, L_x] \neq 0$. The solution to the apparent contradiction lies in the fact that the generalised version of the uncertainty principle reads
\begin{equation}
    \Delta A \Delta B \geq \frac{1}{2} |\braket{D}|,
\end{equation}
where
\begin{equation}
    (\Delta A)^2 = \int \psi^* (A - \braket{A} )^2 \psi \: dt.
\end{equation}
Since the equation is valid for $ \braket{L_y} = 0$, there is no contradiction.
}

A consequence related to that, although came before this derivation, is the concept of indistinguishability of identical particles in quantum mechanics. The concept of identical particles is not unique to quantum mechanics. For instance, John Dalton already assumed that the particles comprising the elements were identical \cite{daltontheory}. But the concept of indistinguishability was something new. In classical physics, even if the particles are identical, they have distinct trajectories defined by their position and momentum. Yet in quantum physics, this cannot be the case, as it is seen by the uncertainty relation. However, the uncertainty relation was not the only means to see this fact, as Dirac had already published a paper on the now so-called Fermi-Dirac distribution of identical fermions in 1926 (about the same time but independently from Fermi). The idea that led Dirac to the indistinguishability is as follows (Dirac, 1926, p.667; \cite{dirac1926} in references):
\begin{quote}
    Consider now a system that contains two or more similar particles, say, for definiteness, an atom with two electrons. Denote by $(mn)$ that state of the atom in which one electron is in an orbit labelled $m$, and the other in the orbit $n$. The question arises whether the two states $(mn)$ and $(nm)$, which are physically indistinguishable as they differ only by the interchange of the two electrons, are to be counted as two different states or as only one state, i.e., do they give rise to two rows and columns in the matrices or to only one? If the first alternative is right, then the theory would enable one to calculate the intensities due to the two transitions $(mn) \to (m'n')$ and $(mn) \to (n'm')$ separately, as the amplitude corresponding to either would be given by a definite element in the matrix representing the total polarisation. The two transitions are, however, physically indistinguishable, and only the sum of the intensities for the two together could be determined experimentally. Hence, in order to keep the essential characteristic of the theory that it shall enable one to calculate only observable quantities, one must adopt the second alternative that $(mn)$ and $(nm)$ count as only one state.
\end{quote}
It is not necessary to point out that these ideas led to philosophical discussions. There was a big controversy between these seemingly contradictory concepts in terms of attaining them to reality at the time (and probably to a greater extent still). The discussion revolved around two keypoints, which may be called the singularities of the philosophical discussions since they were the centers of attractions of conceptual discussions and the origin of the problems themselves. These "singularities" may be summarised more or less as follows:
\begin{enumerate}
    \item The whatness of measurement: What really is measurement in quantum mechanical terms? Is there a point of seperation after which we make a clear distinction between the quantum system and the measurement apparatus?\footnote{It is interesting to see that these questions are peculiarly similar to subject-object problems in philosophy \cite{Bell1973}. See also \cite{Jammer1966}, ch. 7.2.}
    \item The questions on the ontological status of things: Is quantum theory an epistomological theory, that is, a theory which describes nature up to our capabilities of perception as Heisenberg argued \cite{heisenberg1955}; or does it really describe things as it is, and that is how nature really works in the most fundamental level? 
\end{enumerate}
A comprehensive discussions on these matters are clearly out of the scope of this text. However, a term related to these questions and highly relevant to the development of quantum mechanics must be mentioned here: Complementarity. In essence, the idea, which is primarily attributed to Bohr \cite{bohr1928quantum}, states that the nature in its most fundamental level acts in phenomenologically mutually exclusive concepts, such as waves and particles, measurement apparati and systems, or causality and definability of a state (since the act of observation, which is a must that comes before definition, destroys the causal evolution of the system ). More precisely, in Bohr's words (Bohr, 1928, p.580):
\begin{quote}
    On one hand, the definition of the state of
a. physical system, as ordinarily understood,
claims the elimination of all external disturbances.
But in that case, according to the quantum
postulate, any observation will be impossible,
and, above all, the concepts of space and time
lose their immediate sense. On the other hand,
if in order to make observation possible we permit certain interactions with suitable agencies
of measurement, not belonging to the system,
an unambiguous definition of the state of the
system is naturally no longer possible, and
there can be no question of causality in the
ordinary sense of the word. The very nature of the
quantum theory thus forces us to regard the spacetime co-ordination and the claim of causality, the
union of which characterises the classical theories,
as complementary but exclusive features of the
description, symbolising the idealisation of observation and definition, respectively.
\end{quote}
The duels of Einstein and Bohr on these topics were especially illuminating, although generally to Bohr's advantage. One of these discussions led to the EPR paradox \cite{einstein1935can}, Bell inequalities \cite{bell1964}, and finally, by the observation of the violation of inequalities \cite{PhysRevLett.49.91}, to the existence of non-local phenomena in nature (to the advantage of Bohr again).

One of these mental duels is especially interesting, for Bohr's use of general relativity to rule out Einstein's argument \cite{bohr1949discussion}. The argument was as follows: Consider a box filled with radiation. This box has a shutter that operates on clockwork, to open and close for an arbitrarily short time to let a photon escape through the hole. Now, if we measure the mass of the box before and after the event, we can measure the energy of the photon by
\begin{equation}
    E = m c^2,
\end{equation}
and thus violate the energy-time uncertainty relation,
\begin{equation}
    \Delta E \Delta t \gtrsim h.
\end{equation}
Bohr's response was on the following lines: The measurement of the mass difference can be made by balancing the box with a spring and a pointer with an index to read the mass. After the shutter is closed again, an additional weight $m$ can be added to the box to measure the mass difference due to the escaped photon to an accuracy $\Delta m$, which is connected to the accuracy of the position measurement $\Delta q$ of the pointer on the index. Since adding the additional mass will impart an impulse $p$ to the box (and the pointer), which is subjected to the error $\Delta p$ given by
\begin{equation}
    \Delta p \Delta q \gtrsim h.
\end{equation}
The error of the momentum measurement is also bounded by the error of the impulse given by the gravitation as
\begin{equation}
    \Delta p \approx \frac{h}{\Delta q} \leq Tg\Delta m,
\end{equation}
where $T$ is the duration of the balancing procedure, and $g$ is the gravitational constant. But the ticks of the clocks will change throughout the process according to the gravitational redshift formula by
\begin{equation}
    \frac{\Delta T}{T} = \frac{g}{c^2}\Delta q,
\end{equation}
for a displacement $\Delta q$ in the direction of the gravitational field. Substituting for $T$ in the first equation, we get
\begin{equation}
    \Delta T c^2\Delta m = \Delta T \Delta E \gtrsim h.
\end{equation}

\section{Conclusion}
That brings us to the end of our account of the early development of the quantum theory. When we look back, we realise that all the ideas discussed here are actually just derivatives of the concept of "quantum". Planck introduced the concept as a mathematical convenience (which was proven to be a necessity by Poincaré about a decade later). Especially through the discussions and applications of the theory of statistical mechanics (to the calculations of which the concept of quanta brought great convenience as well as new possibilities of thinking, as it did for Planck), the concept had made its place in the minds of the physicist of the time in two decades. The main tools used for calculation were the correspondence principle, in any form, and the quantum of action principle, both of which had a significant place in the fundamentals of matrix mechanics. The correspondence principle is basically a statement which demands you to replace continuity with discreteness, somewhere and somehow. On the other hand, the quantum of action principle directly dictates a quantisation principle. In wave picture, these principles are replaced with boundary conditions on the wave function. That seems too good to be true, and it actually is due to the ontologically suspicious status of the wave function. Most of the time, it cannot be described in our 3-dimensional space, and it also collapses, which is equally indescribable as the other principles. Therefore, collapse is the place where the concept of "quantisation" comes into play, in the sense that we have to empirically assign rules, as it is done in the particle picture, or suppose other things to get rid of it, such as infinite parallel universes \cite{Everett1957} or pilot waves \cite{Bohm1952, Bohm1952II}, which are at best indirectly observable. All in all, maybe the property of "discreteness" (or put in another word, "separability") is somehow equivalent to these seemingly derivative concepts.

Without a doubt, in a work, one of the main aims of which is to provide a better grasp on the fundamental ideas of the quantum theory, not mentioning Pauli's work on spin \cite{Pauli1925}, which is a new degree of freedom unique to the quantum realm; Dirac equation \cite{dirac1928quantum}, which includes the spin degree of freedom as a result of the reconciliation of special theory of relativity and quantum theory; the works of von Neumann with the goal of placing the quantum theory on a mathematically firm ground and understanding the collapse \cite{vonneumann1932mathematische}; (although mentioned) Bell inequalities \cite{bell1964}, the observed violation of which led us to the fact that quantum theory is, in fact, non-local, and thus closed the way to the local hidden variable theories; can be considered a defect. Presumably, for further reading on the topics, the references of the paper, and especially the work of Jammer \cite{Jammer1966}, is a good starting point.

\appendix
\section{Einstein's First Paper on Brownian Motion}
Although not directly related to the quantum theory, since the paper on Brownian motion has historical significance concerning the discrete nature of matter, it seemed relevant to discuss it here \cite{Einstein1956}.

In the first two sections of the paper, Einstein assumes a solvent of volume $V$ and a solute which are separated by a partition $V*$ impermeable to the solute. Then there will be an osmotic pressure $PV*=zRT$ for $z$ gram-molecules on the partition. Now if you assume there are also $n$ suspended particles in the volume, (back then) classical thermodynamics predicts no pressure by these particles on the partition. But Einstein suggests that such particles are differentiated from the solute by only their dimensions and there is no apparent reason for them not to apply pressure on the partition by the thermokinetic movements. Then he proceeds to show mathematically that the molecular-kinetic theory does predict such a pressure of the solute as $PV*=(n/N_A)RT$.

In section three, Einstein derives the diffusion predicted by the classical theory, and in section four, he proceeds to show that molecular theory also predicts that. At the start of the section, he makes two assumptions
\begin{quote}
    \begin{enumerate}
        \item The motions of the particles are independent of one another.
        \item The subsequent motion of individual particles occurring in a time $\tau$ is independent of one another.\footnote{This is considered a weak point in Einstein's argument since the subsequent motions might really well be dependent by the causal relation of initial and final states of movement. A derivation without this presupposition is carried out later, and the formula reached at the end was the same as Einstein's in sufficiently large time intervals: $$\overline{x^2}=2D(t-mB-e^{-t/mB}) \longrightarrow 2Dt$$ where $m$ indicates the mass of the particle and $B$ mobility.}
    \end{enumerate}
\end{quote}
Now, suppose there is an even function $\phi(\Delta)$ representing the probability of a change in the position of the particle from $x$ to $x+\Delta$ (for simplicity, it is studied in 1-dimension denoted $x$).\footnote{The function is even since there is no reason for the probability of a change of $-\Delta$ and $+\Delta$ to have different values.} Let $\eta(x,t)$ represent the number density of the particles per unit volume. By definition, we have
\begin{equation}
    \eta(x,t+\tau)= \int_{\Delta=-\infty}^{\Delta=+\infty} \eta(x-\Delta,t) \phi(\Delta) d\Delta
\end{equation}
and since $\tau$ is very small compared to observable time,
\begin{equation}
    \eta(x,t + \tau) \approx \eta(x,t) + \tau \frac{\partial \eta}{\partial t}.
\end{equation}
Since $\Delta$ is a small displacement compared to the dimension of the volume that contains the particles, we can expand the function in the integral around $\Delta$ and omit the powers after the third term, and combine the expansions to obtain
\begin{equation}
    \begin{split}
    \eta(x,& t + \tau) \approx \eta(x,t) + \tau \frac{\partial \eta}{\partial t}\\
    &=\int_{-\infty}^{+\infty}\eta(x-\Delta,t)\phi(\Delta) d\Delta \approx \int_{-\infty}^{+\infty} \eta(x,t)-\Delta\frac{\partial \eta}{\partial x} + \frac{\Delta^2}{2} \frac{\partial ^2 \eta}{\partial x^2}\phi(\Delta)d\Delta\\
    &=\eta(x,t) \int_{-\infty}^{+\infty} \phi(\Delta) d\Delta - \frac{\partial \eta}{\partial x} \int_{-\infty}^{+\infty} \Delta \phi(\Delta) d\Delta\\ 
    &+\frac{\partial^2 \eta}{\partial x^2} \int_{-\infty}^{+\infty} \frac{\Delta^2}{2} \phi(\Delta) d\Delta.
    \end{split}
\end{equation}
Since $\phi$ and $\Delta$ are even functions, the corresponding integral vanishes, while for $\int_{-\infty}^{+\infty} \phi(\Delta)d\Delta = 1$, $\eta(x,t)$ terms also vanish:
\begin{equation}
    \tau \frac{\partial \eta}{\partial t} = \frac{\partial^2 \eta}{\partial x^2} \int_{-\infty}^{+\infty} \frac{\Delta^2}{2} \phi(\Delta) d\Delta.
\end{equation}
The integral $\int_{-\infty}^{+\infty} \frac{\Delta^2}{2 \tau} \phi(\Delta) d\Delta = D$ is the diffusivity and
\begin{equation}
    \frac{\partial \eta}{\partial t} = D\frac{\partial^2 \eta}{\partial x^2}
\end{equation}
is the equation of diffusion. Assuming that $n$ particles started from the centre at $t=0$, we have the solution
\begin{equation}
    \eta(x,t)=\frac{n}{\sqrt{4 \pi D t}} e^{\frac{-x^2}{4 D t}},
\end{equation}
which is that of a normal distribution. The mean displacement (or the standard deviation) of the particles is thus
\begin{equation}
    \sqrt{\overline{x^2}}= \sqrt{2Dt}.
\end{equation}
Now, using that $D=\frac{RT}{N_A} \frac{1}{6\pi \kappa R}$, where $\kappa$ is a coefficient of viscosity (dynamic viscosity) and $R$ is the radius of the particles,\footnote{This expression is derived in the paper but not included here.} and then equating it with the result we derived, we get a relation that depends on $T, \kappa$ and $R$, whereby acquire a way to determine $N_A$, $\kappa$ or $R$ by measuring the corresponding quantities:
\begin{equation}
   \frac{\overline{x^2}}{2t} = \frac{RT}{N_A} \frac{1}{6\pi \kappa R}.
\end{equation}

\section{Hamilton-Jacobi Formulation of Classical Mechanics\textsuperscript{$\dagger$}}
\blfootnote{$\dagger$ For more information on these topics, see \cite{Goldstein2002Classical} and \cite{HandFinch1998}.}
Finding the equations of motion is easiest when the coordinates are cyclic. Yet the cyclicity of the coordinates depends not only on the system but also on the chosen coordinate representation of the degrees of freedom. Therefore, a procedure to transform a set of variables between different coordinate systems proves useful.

Let a system be described by the coordinate set $\{q_i, p_i\}$, which obeys Hamilton's equations of motion. After a transformation, the new coordinates may be written as
\begin{equation}
    Q_i = Q_i(q, p, t), \: P_i = P_i(q, p, t).
\end{equation}
Since we expect the transformed coordinates to also obey the equations of motions, for transformed Hamiltonian $K = K(Q,P,t)$, we should have
\begin{equation}
    \begin{split}
        & \dot{Q_i} = \frac{\partial K}{\partial P_i}, \\
        & \dot{P_i} = -\frac{\partial K}{\partial Q_i}.
    \end{split}
\end{equation}
Now, since
\begin{equation}
\begin{split}
    \delta \int^{t_2}_{t_1} L dt & = \delta \int^{t_2}_{t_1} p_i\dot{q_i} - H(q,p,t) dt = 0  \\
    & \Longrightarrow \delta \int^{t_2}_{t_1} P_i\dot{Q_i} - K(Q,P,t) dt = 0.
\end{split}
\end{equation}
where $L$ is the Lagrangian of the system, Therefore,
\begin{equation}
   \lambda[ p_i\dot{q_i} - H(q,p,t) ] =  P_i\dot{Q_i} - K(q,p,t) + \frac{dF}{dt},
\end{equation}
where $\lambda$ is independent of $q,p,t$, and $F$ is any function of the phase space coordinates with continuous second derivatives.\footnote{Notice that
\begin{equation}
    \delta \int_{t_1}^{t_2} \frac{dF}{dt}dt = \delta (F_2 - F_1) = 0,
\end{equation}
where $F_i$ are evaluated at the endpoints for any combination of the coordinates $\{Q,P,q,p\}$, along with $t$, they contain. 
}
The transformations for which $\lambda = 1$ are called canonical transformations. The last term of the transformed Lagrangian can be written as a function of transformed or untransformed coordinates, or as a combination of both. They are especially useful when represented as a combination of old and new coordinates, for then they can act as a bridge between the set of coordinates. These types of functions $F$ are called generating functions of the transformation.

Take, for example, the generating function $F_1 = F_1(q, Q, t)$. We have the Lagrangian in terms of the transformed coordinates as
\begin{equation}
    L = p_i\dot{q_i} - H = P_i\dot{Q_i} - K + \frac{dF_1}{dt} = P_i\dot{Q_i} - K + \frac{\partial F_1}{\partial t} + \frac{\partial F_1}{\partial q_i}\dot{q_i} + \frac{\partial F_i}{\partial Q_i}\dot{Q_i}.
\end{equation}
Therefore,
\begin{equation}
    \begin{split}
        & p_i = \frac{\partial F_1}{\partial q_i}, \\
        &  P_i = -\frac{\partial F_1}{\partial Q_i}, \\
        & K = H + \frac{\partial F_1}{\partial t}.
    \end{split}
\end{equation}
Other kinds of generating functions are
\begin{equation}
    \begin{split}
        & F = F_2(q,P,t) - Q_iP_i, \\
        & F = F_3(p,Q,t) +q_ip_i, \\
        & F = F_4(p,P,t) + q_ip_i - Q_iP_i.
    \end{split}
\end{equation}
These are the Legendre transformations of $F_1$. Note that a suitable generating function does not have to be in the form of the four we listed. In all cases, though, the Hamiltonian is in the form
\begin{equation}
\label{K hamiltonian}
    K = H + \frac{\partial F}{\partial t}.
\end{equation}
If the Hamiltonian is conserved, then there is a coordinate frame for which every coordinate is cyclic.\footnote{Observe
\begin{equation}
\begin{split}
    \frac{dH}{dt} & = \frac{\partial H}{\partial  t} + \frac{\partial H}{\partial  q}\dot{q} + \frac{\partial H}{\partial  p}\dot{p} \\
    & = \frac{\partial H}{\partial  t} + \frac{\partial H}{\partial  q}\frac{\partial H}{\partial  p} - \frac{\partial H}{\partial  p}\frac{\partial H}{\partial  q} \\
    & = \frac{\partial H}{\partial t},
\end{split}
\end{equation}
which is equal to zero if the Hamiltonian has no explicit time dependence.
}
Therefore, we can choose a coordinate system where $K = 0$, for then the transformed coordinates satisfy
\begin{equation}
\label{K = Pdot = 0}
    \begin{split}
        & \frac{\partial K}{\partial P_i} = \dot{Q_i} = 0, \\
        -&\frac{\partial K}{\partial Q_i} = \dot{P_i} = 0.
    \end{split}
\end{equation}
By \eqref{K hamiltonian}, then,
\begin{equation}
    H(q,p,t) + \frac{\partial F}{\partial t} = 0.
\end{equation}
It is convenient to express $F$ in $q_i,
P_i$, that is, as $F_2 = F_2(q_i, P_i, t)$, for which we have
\begin{equation}
\label{p_i = partial F_2}
    p_i = \frac{\partial F_2}{\partial q_i},
\end{equation}
so that we have
\begin{equation}
\label{hamiltons pf}
    H(q_i, \frac{\partial F_2}{\partial q_i}, t) + \frac{\partial F_2}{\partial t} = 0,
\end{equation}
which is known as the Hamilton-Jacobi equation. The generating function $F_2$ is denoted as $S$ and called Hamilton's principle function. The total time derivative of it reads
\begin{equation}
    \frac{dS}{dt} = \frac{\partial S}{\partial q_i}\dot{q_i} + \frac{\partial S}{\partial P_i}\dot{P_i} + \frac{\partial S}{\partial t} = p_i\dot{q_i} - H = L,
\end{equation}
by equations \eqref{K = Pdot = 0}, \eqref{p_i = partial F_2}, and \eqref{hamiltons pf}. Therefore,
\begin{equation}
    S = \int L dt + const.
\end{equation}
Yet, since the coordinates are not known as functions of time before solving the problem, we cannot just integrate $L$ and find $S$. When the Hamiltonian does not have an explicit time dependence, $S$ can be written as
\begin{equation}
    S(q, P, t) = W(q, P) - Et,
\end{equation}
where $W$ is called Hamilton's characteristic function. In this case, Hamilton-Jacobi equation takes the special form,
\begin{equation}
    H\left( q_1,..,q_n; \frac{\partial W}{\partial q_1},\dots, \frac{\partial W}{\partial q_n} \right) = E.
\end{equation}
It is interesting to see that, since
\begin{equation}
    \frac{dW}{dt} = \frac{\partial W}{\partial q_i}\dot{q_i} = p_i \dot{q_i},
\end{equation}
we have
\begin{equation}
    W = \int p_i \dot{q_i}dt = \int p_i dq_i,
\end{equation}
which is the action.

\subsection{Separable Systems}
It is customary to express the constant momenta $P_i$ as $\alpha_i$. A coordinate $q_j$ is said to be separable if the principle function can be set in the form
\begin{equation}
\begin{split}
         S(q_1,..,&q_k,\dots,q_n; \alpha_1,..,\alpha_n;t) \\
         & = S_k(q_k;\alpha_1,\dots,\alpha_n;t) + S'(q_1,..,q_{k-1},q_{k+1},\dots,q_n; \alpha_1,..,\alpha_n;t).
\end{split}
\end{equation}
Hamilton's principal function $S$ is said to be completely separable if it can be cast into the form,
\begin{equation}
\label{seperable S}
    S(q_1,\dots,q_n; \alpha_1,\dots,\alpha_n;t) = \sum_{i = 1}^n W_i(q_i;\alpha_1,..,\alpha_n) - Et.
\end{equation}
If we also have a Hamiltonian, such as
\begin{equation}
    H = \sum_{i=1}^n H_i \left( q_i, \frac{\partial W_i}{\partial q_i} \right),
\end{equation}
then the Hamilton-Jacobi equation can be separated into $n$ equations as
\begin{equation}
    H\left( q_i; \frac{\partial W_i}{\partial q_i} \right) = E_i.
\end{equation}
For instance, consider a system with a Hamiltonian which is time-independent and quadratic in momenta, with the potential energy defined by $V = V(q_1,\dots,q_n)$, as
\begin{equation}
    H = \sum_{i=1}^{n} \frac{1}{2m_i} p_i^2 + V(q_1, q_2, \dots, q_n).
\end{equation}
Substituting $p_i = \partial S / \partial q_i$ into the HJE, we get
\begin{equation}
    \sum_{i=1}^{n} \frac{1}{2m_i} \left( \frac{\partial S}{\partial q_i} \right)^2 + V(q_1, q_2, \dots, q_n) = E.
\end{equation}
Assuming that the potential energy is additively separable as
\begin{equation}
    V(q_1, q_2, \dots, q_n) = V_1(q_1) + V_2(q_2) + \dots + V_n(q_n),
\end{equation}
we can assume a completely separable solution for $S$ in the form \eqref{seperable S}. Substituting this into the HJE gives
\begin{equation}
    \sum_{i=1}^{n} \frac{1}{2m_i} \left( \frac{dS_i}{dq_i} \right)^2 +  V_i(q_i) = E.
\end{equation}
Since each term depends only on a single coordinate $q_i$, the equation completely separates into individual equations,
\begin{equation}
\label{HJE for Vxyz}
    \frac{1}{2m_i} \left( \frac{dS_i}{dq_i} \right)^2 + V_i(q_i) = E_i,
\end{equation}
where the total energy is a sum of individual contributions,
\begin{equation}
    E = \sum_{i} E_i.
\end{equation}

\subsection{Action-Angle Variables}
For systems in periodic motion, sometimes the frequency is more relevant to know than the whole solution in terms of the coordinates. In this case, the action-angle variables prove useful, which is the topic of this subsection.

For a system in periodic motion, for which the energy is conserved and the characteristic function is separable, such that
\begin{equation}
    H\left( q_1, \dots, q_n  \ , \frac{\partial W}{\partial q_1}, \dots, \frac{\partial W}{\partial q_n} \right) = \alpha_1, \: W=\sum_i W_i(q_i, \alpha_i).
\end{equation}
The action variables are defined as
\begin{equation}
    J = \oint p_i dq_i = \oint \frac{\partial W_i}{\partial q_i} dq_i.
\end{equation}
Since the $q_i$ dependence is integrated away, $J_i$ are functions of only the constants $\alpha_i$ as $J_i = J_i(\alpha_1, \dots, \alpha_n)$. Assuming that the transformation $(\alpha_1, \dots, \alpha_n) \mapsto (J_1, \dots, J_n) $ is invertible,\footnote{For most integrable systems, we can presume that there are coordinate frames that satisfy the global invertibility conditions. For the conditions and the proof, see \cite{gordon1980diffeomorphisms} and \cite{ohkita2024simple}.} we can write $\alpha_i = \alpha_i(J_1, \dots, J_n)$. Therefore, the characteristic function and the Hamiltonian can be cast into the form
\begin{equation}
    K = K(J) = \alpha_1(J_1, \dots, J_n), \: W=\sum_i W_i(q_i, J_i).
\end{equation}
The angle variable is defined, as usual, by
\begin{equation}
    \theta_i = \frac{\partial W}{\partial J_i}.
\end{equation}
Thus, we have two kinds of canonical variables, $\theta$ and $J$, such that the transformed Hamiltonian depends only on one, as $K = K(J)$, so $\theta$ is cyclic. Since $K$ is a constant of motion (by assumption) and depends only on $J$, $J$ must also be a constant of motion since
\begin{equation}
    \frac{dK}{dt} = \sum_i\frac{\partial K}{\partial J_i}\frac{dJ_i}{dt} = 0 \Longrightarrow \frac{dJ_i}{dt} = 0.
\end{equation}
Therefore, the Hamilton's equation for that system should be in the form
\begin{equation}
    \dot{J_i} = -\frac{\partial K}{\partial \theta_i} = 0, \: \dot{\theta_i} = \frac{\partial K}{\partial J_i} = \omega_i = const.,
\end{equation}
where the reason for the second equality is that since $K = K(J)$, and that $J$ is a constant of motion, then $\dot{\theta}$ can either be zero or another arbitrary constant.

Now, consider a change in $\theta$ in a complete cycle as
\begin{equation}
    \Delta \theta_i = \oint \frac{\partial \theta_i}{\partial q_i} dq_i = \oint \frac{\partial^2 W }{\partial q_i \partial J_i} dq_i = \frac{d}{dJ_i} \oint \frac{\partial W_i}{\partial q_i}dq_i = \frac{dJ_i}{dJ_i} = 1, 
\end{equation}
that is, over a period, the angle variable changes by a unit. Then, since we have
\begin{equation}
    \dot{\theta_i} = \omega_i \Longrightarrow \theta_i = \omega t + c_i,
\end{equation}
where $\omega_i$ and $c_i$ are constants, we should also have
\begin{equation}
    \Delta \theta_i = \omega_i \tau_i = 1 \Longrightarrow \omega_i = 1/\tau_i,
\end{equation}
where $\tau_i$ is the period, and $\omega_i = \partial_{J_i} K$ is the frequency associated with the motion in $q_i$.

\small \textbf{Statements and Declarations}

\footnotesize{ \textbf{Competing Interests}  
The author declares that there are no financial or non-financial competing interests related to this work.

\textbf{Funding}  
No funding was received for conducting this study.

\textbf{Data Availability} No datasets were generated or analyzed during the current study.
}

\newpage
\printbibliography

@article{Einstein1905,
  author = {Albert Einstein},
  title = {Über einen die Erzeugung und Verwandlung des Lichtes betreffenden heuristischen Gesichtspunkt},
  journal = {Annalen der Physik},
  volume = {322},
  number = {6},
  pages = {132--148},
  year = {1905},
  doi = {10.1002/andp.19053220607}
}

@article{Planck1901,
  author = {Planck, Max},
  title = {Ueber das Gesetz der Energieverteilung im Normalspectrum},
  journal = {Annalen der Physik},
  volume = {309},
  number = {3},
  pages = {553-563},
  year = {1901},
  doi = {10.1002/andp.19013090310}
}

@article{KamerlinghOnnesEhrenfest1914,
  author  = {Ehrenfest, Paul and Onnes, Heike Kamerlingh},
  title   = {XXXIII. Simplified deduction of the formula from the theory of combinations which Planck uses as the basis of his radiation theory},
  journal = {The London, Edinburgh, and Dublin Philosophical Magazine and Journal of Science},
  volume  = {29},
  number  = {170},
  pages   = {297--301},
  year    = {1915},
  doi     = {10.1080/14786440208635308},
  url     = {https://doi.org/10.1080/14786440208635308}
}

@inproceedings{Planck1913,
  author    = {Max Planck},
  title     = {La Loi du Rayonnement Noir et l’Hypothèse des Quantités Élémentaires d’Action},
  booktitle = {La théorie du rayonnement et les quanta: rapports et discussions de la réunion tenue à Bruxelles du 30 octobre au 3 novembre 1911},
  year      = {1912},
  pages     = {93 -- 114},
  publisher = {Gauthier-Villars},
  address   = {Paris},
  url       = {http://www.solvayinstitutes.be/pdf/Proceedings_Physics/1911.pdf}
}

@book{Jammer1966,
  author    = {Jammer, Max},
  title     = {The Conceptual Development of Quantum Mechanics},
  edition   = {2nd ed., repr.},
  publisher = {Tomash Publishers and American Institute of Physics},
  year      = {1989},
}

@article{BornJordan1925,
  author    = {Max Born and Pascual Jordan},
  title     = {Zur Quantenmechanik},
  journal   = {Zeitschrift für Physik},
  volume    = {34},
  pages     = {858-888},
  year      = {1925},
  doi       = {10.1007/BF01328531}
}

@article{BornHeisenbergJordan1926,
  author    = {Max Born and Werner Heisenberg and Pascual Jordan},
  title     = {Zur Quantenmechanik II},
  journal   = {Zeitschrift für Physik},
  volume    = {35},
  pages     = {557-615},
  year      = {1926},
  doi       = {10.1007/BF01379806}
}

@article{Heisenberg1925,
  author    = {Werner Heisenberg},
  title     = {Über quantentheoretische Umdeutung kinematischer und mechanischer Beziehungen},
  journal   = {Zeitschrift für Physik},
  volume    = {33},
  pages     = {879-893},
  year      = {1925},
  doi       = {10.1007/BF01328377}
}

@article{BornWiener1926,
  author    = {Max Born and Norbert Wiener},
  title     = {Eine neue Formulierung der Quantengesetze für periodische und nicht periodische Vorgänge},
  journal   = {Zeitschrift für Physik},
  volume    = {36},
  pages     = {174-187},
  year      = {1926},
  doi       = {10.1007/BF01382261}
}

@article{Dirac1925,
  author    = {Paul A. M. Dirac},
  title     = {The Fundamental Equations of Quantum Mechanics},
  journal   = {Proceedings of the Royal Society of London. Series A, Containing Papers of a Mathematical and Physical Character},
  volume    = {109},
  number    = {752},
  pages     = {642-653},
  year      = {1925},
  doi       = {10.1098/rspa.1925.0150}
}

@incollection{Einstein1956,
  author    = {Albert Einstein},
  title     = {On the Movement of Small Particles Suspended in a Stationary Liquid Demanded by the Molecular-Kinetic Theory of Heat},
  booktitle = {Investigations on the Theory of the Brownian Movement},
  editor    = {R. Fürth},
  translator = {A. D. Cowper},
  publisher = {Dover Publications},
  year      = {1956},
  pages     = {1--18},
  address   = {New York}
}

@article{deBroglie1924,
  author    = {Louis de Broglie},
  title     = {A Tentative Theory of Light Quanta},
  journal   = {Philosophical Magazine},
  volume    = {47},
  number    = {278},
  pages     = {446-458},
  year      = {1924},
  doi       = {10.1080/14786442408634378}
}

@article{Schrodinger1926Undulatory,
  author    = {Erwin Schrödinger},
  title     = {An Undulatory Theory of the Mechanics of Atoms and Molecules},
  journal   = {Physical Review},
  volume    = {28},
  pages     = {1049-1070},
  year      = {1926},
  doi       = {10.1103/PhysRev.28.1049}
}

@article{Schrodinger1926Eigenwert1,
  author    = {Erwin Schrödinger},
  title     = {Quantisierung als Eigenwertproblem. Erste Mitteilung},
  journal   = {Annalen der Physik},
  volume    = {384},
  number = {4},
  pages     = {361-376},
  year      = {1926},
  doi       = {10.1002/andp.19263840404}
}

@article{Schrodinger1926Eigenwert2,
  author    = {Erwin Schrödinger},
  title     = {Quantisierung als Eigenwertproblem. Zweite Mitteilung},
  journal   = {Annalen der Physik},
  volume    = {384},
  number = {6},
  pages     = {489-527},
  year      = {1926},
  doi       = {10.1002/andp.19263840602}
}

@article{Schrodinger1926Eigenwert3,
  author    = {Erwin Schrödinger},
  title     = {Quantisierung als Eigenwertproblem. Dritte Mitteilung},
  journal   = {Annalen der Physik},
  volume    = {385},
  number = {13},
  pages     = {437-490},
  year      = {1926},
  doi       = {10.1002/andp.19263851302}
}

@article{Schrodinger1926Eigenwert4,
  author    = {Erwin Schrödinger},
  title     = {Quantisierung als Eigenwertproblem. Vierte Mitteilung},
  journal   = {Annalen der Physik},
  volume    = {386},
  number = {18},
  pages     = {109-139},
  year      = {1926},
  doi       = {10.1002/andp.19263861802}
}

@book{Goldstein2002Classical,
  author    = {Herbert Goldstein and Charles Poole and John Safko},
  title     = {Classical Mechanics},
  edition   = {3rd},
  publisher = {Addison-Wesley},
  year      = {2002},
  %isbn      = {978-0201657029}
}

@article{Hamilton1833,
  author  = {Hamilton, William Rowan},
  title   = {On a General Method of Expressing the Paths of Light, and of the Planets, by the Coefficients of a Characteristic Function},
  journal = {Dublin University Review and Quarterly Magazine},
  volume  = {1},
  pages   = {795--826},
  year    = {1833},
  url     = {https://www.maths.tcd.ie/pub/HistMath/People/Hamilton/LightPlanets/LightPlanets.html}
}

@article{JoasLehner2007,
  author    = {Christian Joas and Christoph Lehner},
  title     = {The classical roots of wave mechanics: Schrödinger’s transformation of the optical-mechanical analogy},
  journal   = {Studies in History and Philosophy of Science Part B: Studies in History and Philosophy of Modern Physics},
  volume    = {40},
  number    = {4},
  pages     = {338-351},
  year      = {2009},
  doi       = {10.1016/j.shpsb.2009.06.007}
}

@article{Einstein_1924,
  author = {Albert Einstein},
  title = {Quantentheorie des einatomigen idealen Gases},
  journal = {Sitzungsberichte der Preußischen Akademie der Wissenschaften, Physikalisch-mathematische Klasse},
  year = {1924},
  pages = {261-267},
  doi = {10.1002/3527608958.ch27}
}

@inbook{einstein1925,
author = {Einstein, A.},
publisher = {John Wiley and Sons, Ltd},
title = {Quantentheorie des einatomigen idealen Gases. Zweite Abhandlung},
booktitle = {Albert Einstein: Akademie‐Vorträge},
pages = {245-257},
doi = {https://doi.org/10.1002/3527608958.ch28},
year = {1925},
}

@book{sommerfeld1921,
  author = {Arnold Sommerfeld},
  title = {Atombau und Spektrallinien},
  year = {1929},
  publisher = {Friedrich Vieweg und Sohn},
  address = {Braunschweig},
  edition = {1st},
  url = {https://archive.org/details/atombauundspektr00somm},
}

@article{ip2013mystery,
  author    = {Pui Him Ip},
  title     = {The Mystery Behind Schrödinger's First Communication: A Non-Historical Study on the Variational Approach and Its Implications},
  journal   = {PhilSci Archive},
  year      = {2013},
  url       = {https://philsci-archive.pitt.edu/9839/},
}

@article{bohr1913,
  author = {Niels Bohr},
  title = {On the Constitution of Atoms and Molecules},
  journal = {Philosophical Magazine},
  volume = {26},
  number = {151},
  pages = {1--25},
  year = {1913},
  doi = {10.1080/14786441308634955}
}

@article{born1926a,
  author = {Max Born},
  title = {Zur Quantenmechanik der Stoßvorgänge},
  journal = {Zeitschrift für Physik},
  volume = {37},
  pages = {863--867},
  year = {1926},
  doi = {10.1007/BF01397477}
}

@article{born1926b,
  author = {Max Born},
  title = {Quantenmechanik der Stoßvorgänge},
  journal = {Zeitschrift für Physik},
  volume = {38},
  pages = {803--827},
  year = {1926},
  doi = {10.1007/BF01397184}
}

@techreport{heisenberg1983,
  author = {Heisenberg, Werner},
  title = {The Actual Content of Quantum Theoretical Kinematics and Mechanics},
  institution = {NASA Technical Memorandum},
  number = {NASA-TM-77379},
  year = {1983},
  url = {https://ntrs.nasa.gov/citations/19840008978}
}

@article{bell1964,
  title = {On the Einstein Podolsky Rosen paradox},
  author = {Bell, J. S.},
  journal = {Physics Physique Fizika},
  volume = {1},
  number = {3},
  pages = {195--200},
  numpages = {6},
  year = {1964},
  month = {Nov},
  publisher = {American Physical Society},
  doi = {10.1103/PhysicsPhysiqueFizika.1.195},
  url = {https://link.aps.org/doi/10.1103/PhysicsPhysiqueFizika.1.195}
}

@book{daltontheory,
  author = {John Dalton},
  title = {A New System of Chemical Philosophy},
  publisher = {Cambridge University Press},
  year = {1808},
  url = {https://archive.org/details/newsystemofchemi01daltuoft},
  pages = {141-145},
}

@article{dirac1926,
  author = {Paul A. M. Dirac},
  title = {On the Theory of Quantum Mechanics},
  journal = {Proceedings of the Royal Society of London. Series A, Containing Papers of a Mathematical and Physical Character},
  volume = {112},
  number = {762},
  pages = {661--677},
  year = {1926},
  doi = {10.1098/rspa.1926.0133},
}

@article{lenard1902,
  author = {Philipp Lenard},
  title = {Über die lichtelektrische Wirkung},
  journal = {Annalen der Physik},
  volume = {313},
  number = {5},
  pages = {149--198},
  year = {1902},
  doi = {10.1002/andp.19023130510}
}

@article{rutherford1911,
  author = {Ernest Rutherford},
  title = {The Scattering of $\alpha$ and $\beta$ Particles by Matter and the Structure of the Atom},
  journal = {Philosophical Magazine},
  volume = {21},
  number = {125},
  pages = {669--688},
  year = {1911},
  doi = {10.1080/14786440508637080}
}

@book{bohr1918,
  author    = {Niels Bohr},
  title     = {On the Quantum Theory of Line-Spectra},
  publisher = {Project Gutenberg},
  year      = {2014},
  url       = {https://www.gutenberg.org/ebooks/47167}
}

@article{poincare1921,
  author    = {Henri Poincaré},
  title     = {Sur la théorie des quanta},
  journal   = {Journal de Physique Théorique et Appliquée},
  volume    = {2},
  number    = {1},
  year      = {1912},
  pages     = {5 - 34},
  doi = {10.1051/jphystap:0191200200500}
}

@article{sommerfeld1915,
  author    = {Arnold Sommerfeld},
  title     = {Zur Theorie der Balmerschen Serie},
  journal   = {Sitzungsberichte der Bayerischen Akademie der Wissenschaften, Mathematisch-Physikalische Klasse},
  year      = {1915},
  pages     = {425--458},
  url       = {https://d-nb.info/1345425740/34}
}

@article{sommerfeld1916,
author = {Sommerfeld, A.},
title = {Zur Quantentheorie der Spektrallinien},
journal = {Annalen der Physik},
volume = {356},
number = {17},
pages = {1-94},
doi = {https://doi.org/10.1002/andp.19163561702},
year = {1916}
}

@article{born1924,
  author    = {Max Born},
  title     = {Über Quantenmechanik},
  journal   = {Zeitschrift für Physik},
  volume    = {26},
  year      = {1924},
  pages     = {379--395},
  doi       = {10.1007/BF01327341}
}

@book{HandFinch1998,
  author    = {Louis N. Hand and Janet D. Finch},
  title     = {Analytical Mechanics},
  year      = {1998},
  publisher = {Cambridge University Press},
  address   = {Cambridge, UK}
}

@article{gordon1980diffeomorphisms,
  title={On the Diffeomorphisms of Euclidean Space},
  author={Gordon, William B.},
  journal={The American Mathematical Monthly},
  volume={79},
  number={7},
  pages={755--759},
  year={1972},
  publisher={JSTOR},
  doi = {10.2307/2316266}
}

@article{ohkita2024simple,
  author    = {Shinobu Ohkita and Masaki Tsukamoto},
  title     = {Simple proof of the global inverse function theorem via the Hopf–Rinow theorem},
  journal   = {Proceedings of the Japan Academy, Series A, Mathematical Sciences},
  volume    = {100},
  number    = {3},
  pages     = {17 - 19},
  year      = {2024},
  doi       = {10.3792/pjaa.100.004},
  url       = {https://arxiv.org/pdf/2306.17410}
}

@article{Aitchison2004,
  author    = {Ian J. R. Aitchison and David A. MacManus and Thomas M. Snyder},
  title     = {Understanding Heisenberg's 'magical' paper of July 1925: a new look at the calculational details},
  journal   = {American Journal of Physics},
  volume    = {72},
  number    = {11},
  pages     = {1370--1379},
  year      = {2004},
  doi       = {10.1119/1.1775243}
}

@article{Pauli1925,
  author    = {Pauli, W.},
  title     = {Über den Einfluß der Geschwindigkeitsabhängigkeit der Elektronenmasse auf den Zeemaneffekt},
  journal   = {Zeitschrift für Physik},
  volume    = {31},
  pages     = {373--385},
  year      = {1925},
  doi       = {10.1007/BF02980592},
  url       = {https://link.springer.com/article/10.1007/BF02980592}
}

@article{dirac1928quantum,
  author = {Dirac, P. A. M.},
  title = {The Quantum Theory of the Electron},
  journal = {Proceedings of the Royal Society of London. Series A, Containing Papers of a Mathematical and Physical Character},
  volume = {117},
  number = {778},
  pages = {610--624},
  year = {1928},
  publisher = {The Royal Society},
  doi = {10.1098/rspa.1928.0023}
}

@book{vonneumann1932mathematische,
  author = {von Neumann, John},
  title = {Mathematische Grundlagen der Quantenmechanik},
  year = {1932},
  publisher = {Springer}
}

@book{heisenberg1955,
  author    = {Werner Heisenberg},
  title     = {Das Naturbild der heutigen Physik},
  year      = {1955},
  publisher = {Rowohlt},
  address   = {Hamburg}
}

@incollection{Bell1973,
	author = {John S. Bell},
	booktitle = {The physicist's conception of nature},
	editor = {Jagdish Mehra},
	pages = {687--690},
	publisher = {Reidel},
	title = {Subject and Object},
	year = {1973}
}

@article{bohr1928quantum,
  author    = {Niels Bohr},
  title     = {The Quantum Postulate and the Recent Development of Atomic Theory},
  journal   = {Nature},
  volume    = {121},
  pages     = {580--590},
  year      = {1928},
  doi       = {10.1038/121580a0}
}

@article{einstein1935can,
  author    = {Albert Einstein and Boris Podolsky and Nathan Rosen},
  title     = {Can Quantum-Mechanical Description of Physical Reality Be Considered Complete?},
  journal   = {Physical Review},
  volume    = {47},
  number    = {10},
  pages     = {777--780},
  year      = {1935},
  doi       = {10.1103/PhysRev.47.777}
}

@article{PhysRevLett.49.91,
  title = {Experimental Realization of Einstein-Podolsky-Rosen-Bohm Gedankenexperiment: A New Violation of Bell's Inequalities},
  author = {Aspect, Alain and Grangier, Philippe and Roger, G\'erard},
  journal = {Phys. Rev. Lett.},
  volume = {49},
  number = {2},
  pages = {91--94},
  numpages = {0},
  year = {1982},
  month = {Jul},
  publisher = {American Physical Society},
  doi = {10.1103/PhysRevLett.49.91},
}

@incollection{bohr1949discussion,
  author    = {Niels Bohr},
  title     = {Discussion with Einstein on Epistemological Problems in Atomic Physics},
  booktitle = {Albert Einstein: Philosopher-Scientist},
  editor    = {Paul Arthur Schilpp},
  publisher = {Library of Living Philosophers},
  year      = {1949},
  pages     = {201--241}
}

@article{Kirchhoff1860,
  author = {G. Kirchhoff},
  title = {Über das Verhältnis zwischen dem Emissionsvermögen und dem Absorptionsvermögen der Körper für Wärme und Licht},
  journal = {Annalen der Physik},
  volume = {185},
  number = {2},
  pages = {275-301},
  year = {1860},
  doi = {10.1002/andp.18601850205}
}

@article{Wien1896,
  author = {W. Wien},
  title = {Über die Energieverteilung im Emissionsspectrum eines schwarzen Körpers},
  journal = {Annalen der Physik},
  volume = {294},
  number = {8},
  pages = {662-669},
  year = {1896},
  doi = {10.1002/andp.18962940803}
}

@article{Rayleigh1900,
  author = {Lord Rayleigh},
  title = {Remarks upon the law of complete radiation},
  journal = {Philosophical Magazine},
  volume = {49},
  number = {301},
  pages = {539-540},
  year = {1900},
  doi = {10.1080/14786440009463878}
}

@article{ehrenfest1911,
  author    = {Paul Ehrenfest},
  title     = {Welche Z{\"u}ge der Lichtquantenhypothese spielen in der Theorie der W{\"a}rmestrahlung eine wesentliche Rolle?},
  journal   = {Annalen der Physik},
  volume    = {341},
  number    = {11},
  pages     = {91--118},
  year      = {1911},
  doi       = {10.1002/andp.19113411106}
}

@article{kennard1927,
  author    = {E. H. Kennard},
  title     = {Zur Quantenmechanik einfacher Bewegungstypen},
  journal   = {Zeitschrift f{\"u}r Physik},
  year      = {1927},
  volume    = {44},
  pages     = {326--352},
  doi       = {10.1007/BF01391200}
}

@article{Bohm1952,
  author = {David Bohm},
  title = {A Suggested Interpretation of the Quantum Theory in Terms of "Hidden" Variables. I},
  journal = {Physical Review},
  volume = {85},
  number = {2},
  pages = {166--179},
  year = {1952},
  doi = {10.1103/PhysRev.85.166}
}

@article{Bohm1952II,
  title = {A Suggested Interpretation of the Quantum Theory in Terms of "Hidden" Variables. II},
  author = {Bohm, David},
  journal = {Physical Review},
  volume = {85},
  number = {2},
  pages = {180--193},
  year = {1952},
  publisher = {American Physical Society},
  doi = {10.1103/PhysRev.85.180}
}

@book{Everett1957,
  author = {Everett, Hugh},
  title = {The Many-Worlds Interpretation of Quantum Mechanics},
  year = {1957},
  publisher = {Princeton University},
  note = {PhD Thesis}
}

@article{Bose1924,
  author = {S. N. Bose},
  title = {Planck’s Gesetz und Lichtquantenhypothese},
  journal = {Zeitschrift für Physik},
  volume = {26},
  pages = {178--181},
  year = {1924},
  doi = {10.1007/BF01327326}
}

@incollection{Planck1900,
  author    = {Planck, Max},
  title     = {Über die Verbesserung der Wien’schen Spektralgleichung},
  booktitle = {Von Kirchhoff bis Planck},
  editor    = {Schöpf, Hans-Georg},
  series    = {Reihe Wissenschaft},
  publisher = {Vieweg+Teubner Verlag},
  address   = {Wiesbaden},
  year      = {1900},
  pages     = {175--178},
  note      = {Reprint of the original 1900 article in \emph{Verhandlungen der Deutschen Physikalischen Gesellschaft}}
}

@article{Jeans1905,
  author = {J. H. Jeans},
  title = {On the Laws of Radiation},
  journal = {Philosophical Transactions of the Royal Society of London. Series A, Containing Papers of a Mathematical or Physical Character},
  volume = {76},
  number = {513},
  year = {1905},
  pages = {545--552},
  publisher = {Royal Society},
  doi = {10.1098/rspa.1905.0060}
}

@inproceedings{Bomark2023,
  author       = {Nils-Erik Bomark and Reidun Renstrøm},
  title        = {The Ultraviolet Myth},
  booktitle    = {Proceedings of the European Physical Society Conference on High Energy Physics (EPS-HEP 2023)},
  year         = {2023},
  url          = {https://pos.sissa.it/449/660/}
}
\end{document}